\documentclass[aps,prl,twocolumn,showpacs,nofootinbib,superscriptaddress]{revtex4-2}
\usepackage{amsmath}
\usepackage{amsfonts}
\usepackage{amssymb}
\usepackage{mathtools}
\usepackage{bm}          
\usepackage{dcolumn}     
\usepackage{slashed}     
\usepackage{upgreek}     
\usepackage{bigints}     
\usepackage{graphicx}
\usepackage{epsfig}
\usepackage{pifont}
\usepackage[dvipsnames,table]{xcolor}
\usepackage{colortbl}
\usepackage{adjustbox}
\usepackage[colorlinks=true,linkcolor=blue]{hyperref}%
\usepackage{rotating}
\usepackage{float}
\usepackage[caption=false]{subfig}

\usepackage{array}
\usepackage{multirow}
\usepackage{makecell}
\usepackage{hhline}
\usepackage{diagbox}     

\usepackage{placeins}
\usepackage{textcomp}
\usepackage{blindtext}
\usepackage{url}
\usepackage[utf8]{inputenc}
\usepackage{docs}
\usepackage[toc,page]{appendix}
\usepackage[normalem]{ulem}
\usepackage{lipsum}
\usepackage{chngcntr}
\usepackage[shortlabels]{enumitem}
\usepackage{empheq}

\usepackage[colorlinks=true,linkcolor=blue]{hyperref}
\usepackage{cleveref}
\usepackage[toc,page]{appendix}
\usepackage{tikz}
\usepackage{tikz-feynman}
\usepackage{pifont}
\usetikzlibrary{snakes, arrows}
\tikzfeynmanset{compat=1.1.0}
\tikzset{vertex/.style={circle,draw, minimum size=1.5em}, edge/.style={->,> = latex'}}
\def\Lc{\Lambda_c}
\def\Lb{\Lambda_b}
\def \lb{\Lambda_b}
\def \lc{\Lambda_c}
\def\thl{{\theta_\ell}}
\def\nn{\nonumber}

\newcommand{\bra}[1]{\langle #1|}
\newcommand{\ket}[1]{|#1\rangle}

\begin{document}

\title{ Signatures of Invisible Fermions in $\Lambda_b^0 \to \Lambda_c^+ \ell^- \bar{X}_{\rm inv}$ Decays } 
\author{Shantanu Sahoo}
\affiliation{School of Physics, University of Hyderabad, Hyderabad, Telangana-500046, India}
\affiliation{Department of Physics, Indian Institute of Technology Guwahati, North Guwahati, Assam-781039, India}
\author{Soumitra Nandi}
\affiliation{Department of Physics, Indian Institute of Technology Guwahati, North Guwahati, Assam-781039, India}


\begin{abstract}
We investigate the effects of a massive invisible fermion in $\Lambda_b^0 \to \Lambda_c^+ \ell^- \bar{X}_{\rm inv}$ decays using a model-independent weak effective theory. Through a comprehensive angular analysis, we show that a nonzero invisible-particle mass leaves characteristic imprints on angular observables that can distinguish massless and massive invisible states. We further demonstrate that these observables provide strong discrimination among vector/axial-vector, scalar/pseudoscalar, and tensor interactions, as well as between left- and right-handed quark and lepton current operators. 
\end{abstract}

\maketitle

\paragraph{\underline{Introduction}:}
Semileptonic decays $\Lambda_b^0 \to \Lambda_c^+ \ell^- \bar{X}_{\rm inv}$ provide an important avenue for precision tests of the Standard Model (SM) and the determination of the CKM matrix element $|V_{cb}|$. 
In the SM, the invisible particle is a nearly massless neutrino; however, the missing-energy signature associated with these decays also offers a unique opportunity to test the presence of non-standard invisible states. The missing-mass-squared distributions measured in such semileptonic decay processes exhibit a peak near zero with a finite spread~\cite{Belle:2023bwv,Belle-II:2023okj}, motivating the exploration of scenarios in which the invisible particle is not strictly massless. Light sterile neutrinos and dark-sector fermions provide well-motivated beyond-the-Standard-Model (BSM) realizations of invisible particles \cite{Kolay:2026mgv,Kolay:2026bjm}. Their presence can significantly modify the phase space and angular observables of semileptonic $\Lambda_b$ decays, making them sensitive probes of such scenarios.

In this work, we study the decay $\Lambda_b \to \Lambda_c \ell \chi$ within a model-independent effective field theory framework that allows both left- and right-handed interactions. Particular emphasis is placed on the impact of the invisible-fermion mass on decay kinematics and angular observables. We demonstrate that several angular observables exhibit enhanced sensitivity to $m_\chi$ and can effectively discriminate between different chiral interaction structures, thereby providing a promising probe of light invisible fermions and dark-sector scenarios in flavour physics.

\begin{figure}[t]
	\centering
	\includegraphics[width=0.85\linewidth]{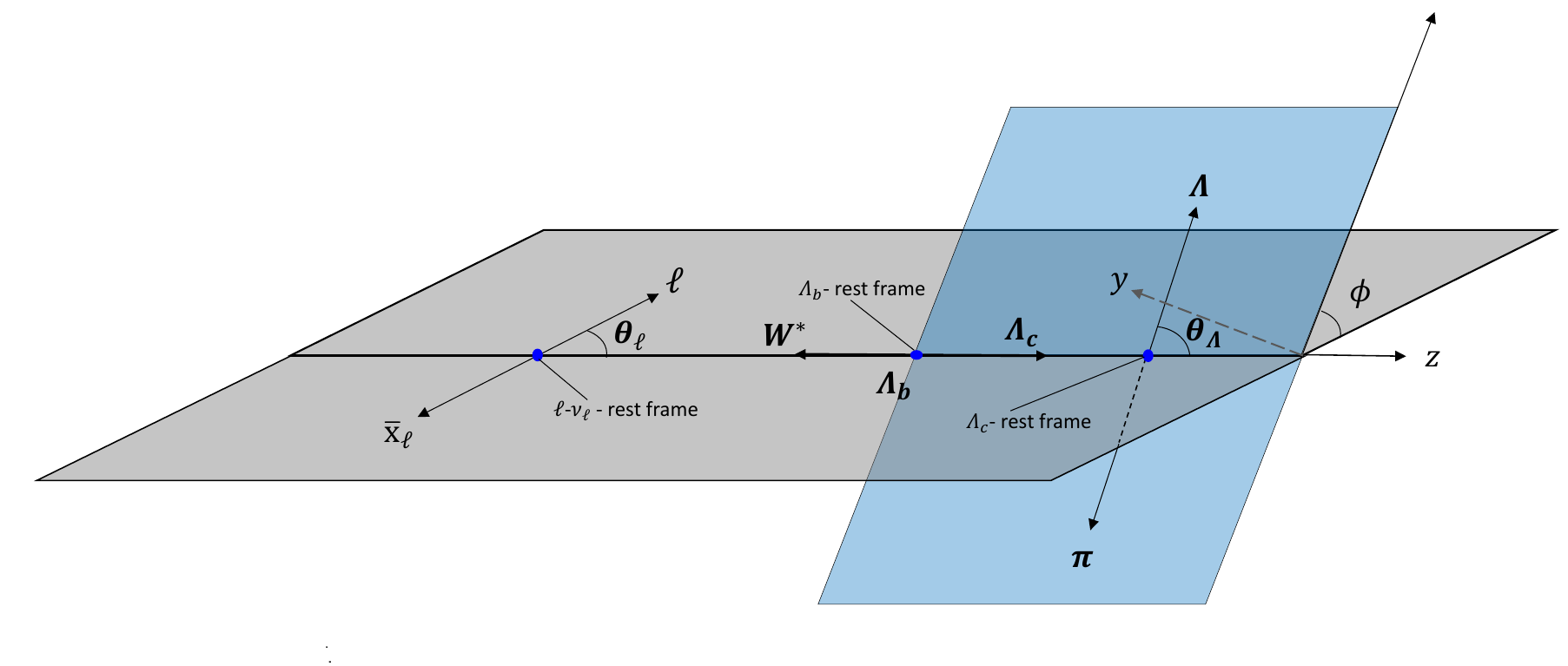}
	\caption{Schematic diagram of the decay plane of $\Lb \to \Lc (\to \Lambda \, \pi) \ell \chi $. }
	\label{fig:deacy_dia}
\end{figure}

 
\paragraph{\underline{Theory Framework}:}
We describe the decay $ \Lb \to \Lc \ell X_{\rm inv}$ within a general weak effective theory framework, where $X_{\rm inv}$ denotes either a SM neutrino or a light invisible fermion $\chi$. The effective Hamiltonian for $b \to c \ell X_{\rm inv}$ (with $\ell = e, \, \mu$) transitions is
\begin{equation}\label{eq:general_Hamiltonian_b2c}
	\mathcal{H}_{\rm eff} = \frac{4 G_F}{\sqrt{2}} V_{cb} \sum_i C_i^\ell \, \mathcal{O}_i^\ell \, ~,
\end{equation}
where we retain the relevant vector, scalar, and tensor operators,
\begin{widetext}
\begin{align}
\label{eq:operators}
 & \mathcal{O}_{V_{1L}}^{\ell,SM} = \left(\bar{c}\gamma_{\mu} P_{L} b \right)\left( \bar{\ell} \gamma^{\mu} P_{L}\nu_{\ell} \right), \,\ \,  
  \mathcal{O}_{V_{1L(R)}}^{\ell} = \left(\bar{c}\gamma_{\mu} P_{L} b \right)\left( \bar{\ell} \gamma^{\mu} P_{L(R)} \chi \right)\,,\ \ \ 
 \mathcal{O}_{V_{2L(R)}}^{\ell}  = \left(\bar{c}\gamma_{\mu} P_{R} b \right)\left( \bar{\ell} \gamma^{\mu} P_{L(R)} \chi \right)\,,  \nn \\ 
 &\mathcal{O}_{S_{1L(R)}}^{\ell} = \left(\bar{c} P_{R}  b \right)\left( \bar{\ell} P_{L(R)} \chi \right)\,,\ \ 
 \mathcal{O}_{S_{2L(R)}}^{\ell} = \left(\bar{c}P_{L} b \right)\left( \bar{\ell} P_{L(R)} \chi \right). \, \ \  
\mathcal{O}_{TL(R)}^{\ell} = \left(\bar{c}\sigma_{\mu\nu} P_{L(R)} b \right)\left( \bar{\ell} \sigma^{\mu \nu } P_{L(R)}) \chi \right)\,. 
\end{align}
\end{widetext}
The SM contribution corresponds to $\mathcal{O}_{V_{1L}}^{\ell,SM}$ with $C_{V_{1L}}^{\ell,SM}=1$, involving left-handed neutrinos. In contrast, the dark fermion $\chi$ can appear with both chiralities, giving rise to additional operator structures. Throughout this work, we restrict to real Wilson coefficients (WCs) and focus on the light leptons $\ell=(e,\mu)$. For $\chi$ to manifest as missing energy, its mass must satisfy
$m_\chi \leq (m_{\Lb} - m_{\Lc} - m_\ell)$. Such operators naturally arise in dark-sector effective theories and simplified models (see, e.g.,~\cite{Aebischer:2022wnl}). In particular, the operators in eq.~\eqref{eq:general_Hamiltonian_b2c} can be matched onto the dimension-6 dark low-energy effective theory (DLEFT) basis \cite{Aebischer:2022wnl}, which extends the SM low-energy basis by including dark-sector states, valid below the electroweak scale with $SU(3)_c \otimes U(1)_{\rm em}$ invariance.
Moreover, ultraviolet-complete frameworks provide a well-motivated origin for these effective interactions. In particular, multicomponent dark matter models with scalar or vector leptoquark mediators~\cite{Belfatto:2021ats}, lepton-portal dark matter scenarios~\cite{Okawa:2020jea, Iguro:2022tmr, Higuchi:2023kbt} and right-handed sterile neutrino models~\cite{Drewes:2016upu, Boyarsky:2018tvu}. These frameworks can generate the operators in eq.~\eqref{eq:general_Hamiltonian_b2c} at tree level with WCs of ${\cal O}(1)$ \cite{Kolay:2026mgv, Kolay:2026bjm}, thereby offering a compelling top-down origin for the effective Hamiltonian considered in this study.
\paragraph{\underline{Angular Observables} :}
We study the angular observables associated with four-fold differential decay distributions of $\Lb \to \Lc(\to \Lambda \pi) \ell X_{\rm inv}$, where $X_{\rm inv}$ can be either a SM neutrino or a light dark-sector fermion $\chi$ with mass $ m_{\chi} $.
Since both escape detection, they appear as missing energy; however, a non-zero $m_\chi$ modifies the available phase space. In the SM, the dilepton invariant mass-squared $q^2 \in \left[m_\ell^2,\,(m_{\Lb} - m_{\Lc})^2\right]$, whereas for a massive invisible particle $q^2$ range shifts to $\left[(m_\ell + m_\chi)^2,\,(m_{\Lb} - m_{\Lc})^2\right]$, leading to characteristic changes in the decay distributions. Following the discussion,
the decay width for $\Lb \to \Lc \ell X_{\rm inv}$ can be written as
{\small
\begin{equation}\label{eq:total_decay_width}
	\Gamma(\Lb \to \Lc \ell X_{\rm inv}) = \Gamma(\Lb \to \Lc \ell \nu) \Bigr|_{SM} + \Gamma(\Lb \to \Lc \ell \chi )\Bigr|_{NP}\,.
\end{equation}
}
The detailed mathematical expressions for the decay widths $\Gamma(\Lb \to \Lc \ell \nu)|_{SM}$ and the respective decay distributions are available in the literature~\cite{Nandi:2024aia, Datta:2017aue}.
The differential decay width for the process $\Gamma(\Lb \to \Lc (\to \Lambda \pi) \ell \chi)$ can be obtained by integrating the four-fold angular distribution which can be written as:
\begin{widetext}
\begin{align}\label{eq:fivefold_decaywidth}
		\frac{d}{ dk^2}\left(\frac{d^4 \Gamma }{d q^2 \, d\cos \theta_{\Lambda} \,  d\cos \theta_{\ell} \, d\phi} \right) = & \frac{\lambda^{1/2}(p_{\lb}^2, p_{\lc}^2, q^2)}{1024 (2\pi)^6 m_{\lb}^3} 
		\frac{\lambda^{1/2}(k^2, p_\Lambda^2, p_\pi^2)}{m_{k}^2}  
		\frac{\lambda^{1/2}(q^2, p_{\ell}^2, p_{\chi}^2)}{q^2} \sum_{\lambda, \lambda_{\ell}} \sum_{\lambda_{\Lc}} \left| \mathcal{M}^{\lambda}_{\lambda_{\Lc}, \lambda_{\ell}}(\lb \to \Lambda \pi \ell \bar{\chi}) \right|^2 \, ,
	\end{align}
\end{widetext}
with the K{\"a}llen $\lambda$ function defined as $\lambda(\alpha,\beta,\gamma) = \alpha^2+\beta^2+\gamma^2-2\alpha\beta - 2\beta\gamma-2\alpha\gamma$. The corresponding SM result can then be recovered by retaining only the contribution from the operator $\mathcal{O}_{V_{1L}}^{\ell,SM}$ and taking the limit of a massless $\chi$. In the above equation $\mathcal{M}^{\lambda}_{\lambda_{\Lc}, \lambda_{\ell}}(\Lb \to \Lambda \pi \ell \chi )$ represents the four-body helicity amplitude and the parameters $\lambda_\ell$, $\lambda_{\Lc}$ and $\lambda_{\chi}$, respectively, represent the helicities of the charged lepton, $\Lc$ baryon and dark sector fermion $\chi$. The calculation details for this helicity amplitude are provided in the appendix~\ref{appsec:helicityamp}. In the above equation,
$\theta_\Lambda,\, \thl, \,$ and $\phi$ are different polar and azimuthal angles, defined in the decay plane of fig.~\ref{fig:deacy_dia}.
The resulting four-fold decay distribution is then obtained as follows:
\begin{widetext}
\begin{align}~\label{eq:total_dist}
	\frac{d^4 \Gamma (\Lb \to \Lc \ell X_{\mathrm{inv}})}{dq^2 \, d \cos\thl \, d\cos \theta_{\Lambda} \, d\phi} =\frac{d^4 \Gamma (\Lb \to \Lc \ell \nu)}{dq^2 \, d \cos\theta_\ell \, d\cos\theta_{\Lambda} \, d\phi}
	+ \frac{d^4 \Gamma (\Lb \to \Lc \ell \chi)}{dq^2 \, d \cos\theta_\ell \, d\cos\theta_{\Lambda} \, d\phi}
    = \dfrac{3}{8 \pi} \times K(q^2, m_{\chi}^2,\theta_\ell,\theta_{\Lambda},\phi) .
\end{align}
where,
\begin{align}~\label{eq:dGamma}
	K(q^2, \cos\theta_\ell, \cos\theta_\Lambda, \phi)
	& =
	\big( K_{1ss} \sin^2\theta_\ell +\, K_{1cc} \cos^2\theta_\ell + K_{1c} \cos\theta_\ell\big) \,\cr
	&  + \big( K_{2ss} \sin^2\theta_\ell +\, K_{2cc} \cos^2\theta_\ell + K_{2c} \cos\theta_\ell\big) \cos\theta_\Lambda
	\cr
	&  + \big( K_{3sc}\sin\theta_\ell \cos\theta_\ell + K_{3s} \sin\theta_\ell\big) \sin\theta_\Lambda \cos\phi\cr
	&  + \big( K_{4sc}\sin\theta_\ell \cos\theta_\ell + K_{4s} \sin\theta_\ell\big) \sin\theta_\Lambda \sin\phi \,.
\end{align}
\end{widetext}
The associated angular coefficients can be extracted as
{\small
\begin{equation}\label{eq:angular_coefficients}
	K_i(q^2, m_{X_{\mathrm{inv}}}^2)
	= K_i(q^2)^{\Lb \to \Lc \ell \nu }
	+ K_i(q^2, m_{\chi}^2)^{\Lb \to \Lc \ell \chi }.
\end{equation}
}
These angular coefficients are directly related to the hadronic helicity amplitudes. It is important to note that the kinematically allowed $q^2$ range differs between the massless neutrino case in the SM and the scenario with a massive dark-sector particle. 
The above-defined angular coefficients can be redefined as $\hat{K}(q^2)$ in which the systematic uncertainties are expected to cancel in the ratio. The normalized angular coefficients are defined as:
\begin{equation}
    \hat{K}_i(q^2)= \frac{K_i(q^2)^{\Lb \to \Lc \ell \nu }
    	+ K_i(q^2, m_{\chi}^2)^{\Lb \to \Lc \ell \chi }}{\frac{d\Gamma(\Lb \to \Lc\ell X_{\mathrm{inv}})}{dq^2}},
    \label{eq:ang_normalised}
\end{equation}
In the above equation, $d\Gamma(\Lb \to \Lc \ell X_{\mathrm{inv}})/dq^2$ will be obtained from  eq.~\eqref{eq:total_dist}. 

The asymmetry observables are defined from eq.~\eqref{eq:dGamma} as:
	\begin{equation}
		A^\ell_\text{FB}(q^2) = \frac{3}{2} \, \frac{K_{1c}(q^2)}{2 K_{1ss}(q^2) + K_{1cc}(q^2)},   
	\end{equation}
    \begin{align}
		A^{\Lambda_c}_\text{FB}(q^2)
		& = \frac{1}{2} \, \frac{2 K_{2ss}(q^2) + K_{2cc}(q^2)}{2 K_{1ss}(q^2) + K_{1cc}(q^2)}. 
	\end{align}
    \begin{align}
		A^{\Lambda_c \ell}_\text{FB}(q^2) & = \frac{3}{4} \, \frac{K_{2c}(q^2)}{2 K_{1ss}(q^2) + K_{1cc}(q^2)},
	\end{align}
    Detailed discussion of these observables is provided in the appendix \ref{appsec:angular_Obs}. 
\paragraph{\underline{Analysis and Results}:}
The helicity amplitudes of the hadronic current depend on the non-perturbative form factors and their $q^2$ shapes. Hence, to get the shape of the decay rate distribution, one needs to know the shape of the corresponding form factors in the whole $q^2$ region. The lattice results for the relevant form factors are available in~\cite{Detmold:2015aaa, Datta:2017aue}. To obtain the $q^2$ shapes of the form factors, we used the $z$-parameterization~\cite{Detmold:2015aaa, Datta:2017aue} read as
\begin{equation}
	f(q^2)=\frac{1}{1-q^2/(m_{\text{pole}}^f)^2}\Big{[}a_{0}+a_{1}z^{f}(q^2) + a_{2}[z^{f}(q^2)]^2\Big{]} \, ,
	\label{eq:zseries_exp}
\end{equation}
where the parameters $a^f_{0}$ and $a^f_{1}$, $a^f_{2}$ are BCL coefficients. The expansion parameter $z^f$ is defined as
\begin{equation}
	z^f(q^2)=\frac{\sqrt{t_+^f-q^2}-\sqrt{t_+^f-t_0}}{\sqrt{t_+^f-q^2}+\sqrt{t_+^f-t_0}} \, ,
\end{equation}
with $t_+^f=(m_\text{pole}^f)^2$ and $t_{0}^f=(m_{\Lb} - m_{\Lc})^2$. The pole masses $m_{\text{pole}}^f$, used in eq.~\ref{eq:zseries_exp} for the respective form factors, are listed together with their associated spin quantum number in the appendix~\ref{appsec:FormFactor_info}. The values of the parameters $a^f_{0}$, $a^f_{1}$ and $a^f_{2}$ together with their covariance matrix are provided in ref.~\cite{Detmold:2015aaa, Datta:2017aue} and are summarized in the appendix~\ref{appsec:FormFactor_info}. We used these inputs to obtain the shapes of the form factors. Using these shapes, we have obtained the decay rate distributions and predicted many other related observables. 
\begin{figure*}[t]
	\centering
	\vspace{-0.1cm}
	\subfloat[]{
		\includegraphics[scale=0.43]{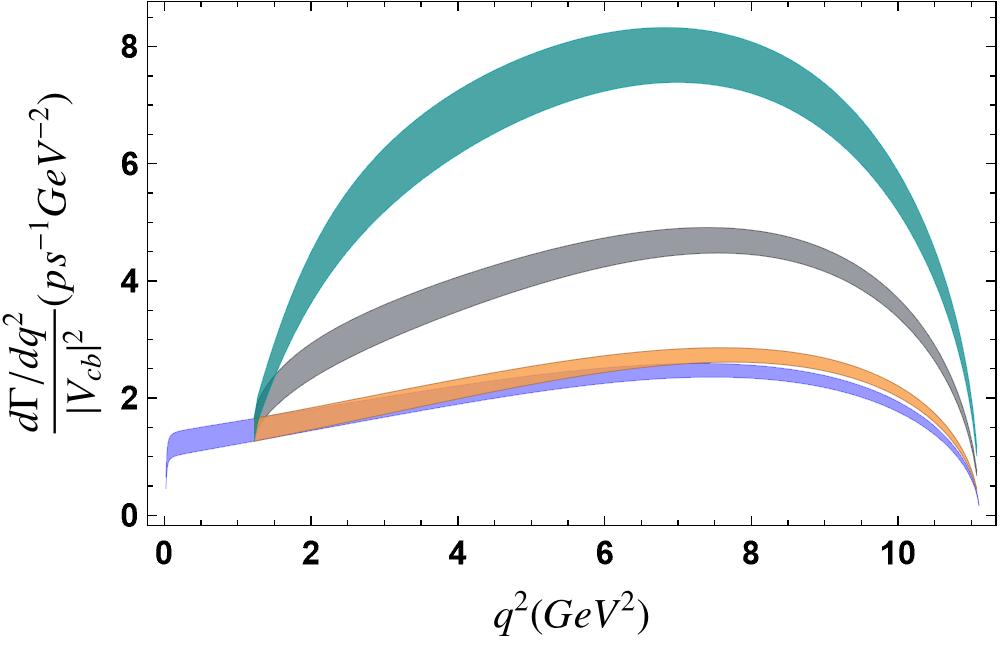}}
	\vspace{-0.1cm}\hspace{0.0002cm}
	\subfloat[]{\includegraphics[scale=0.43]{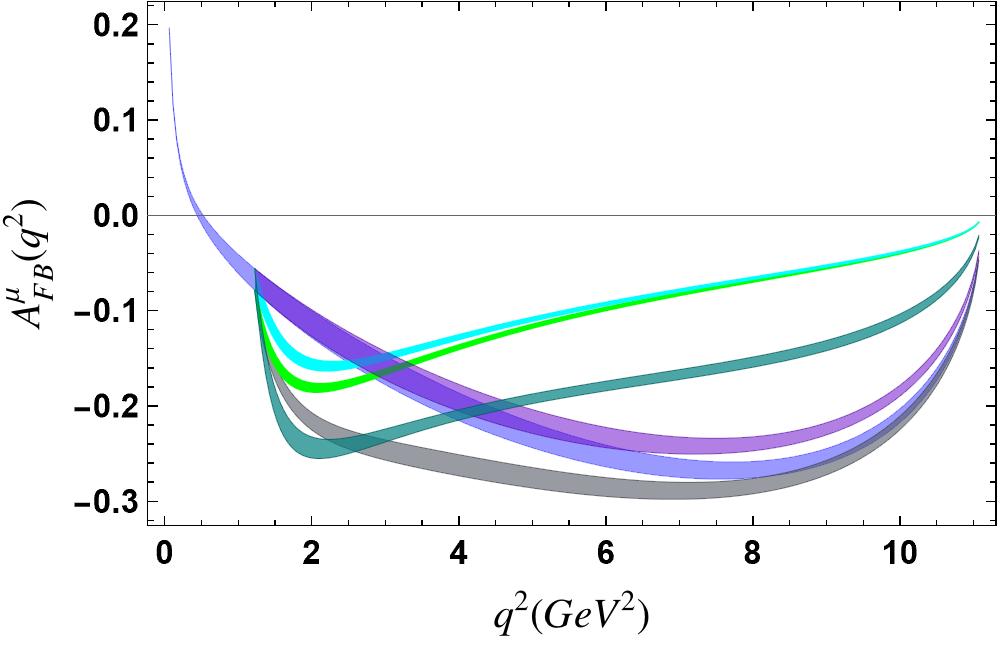}}
	\vspace{-0.1cm}\hspace{0.0002cm}
	\subfloat[]{\includegraphics[scale=0.43]{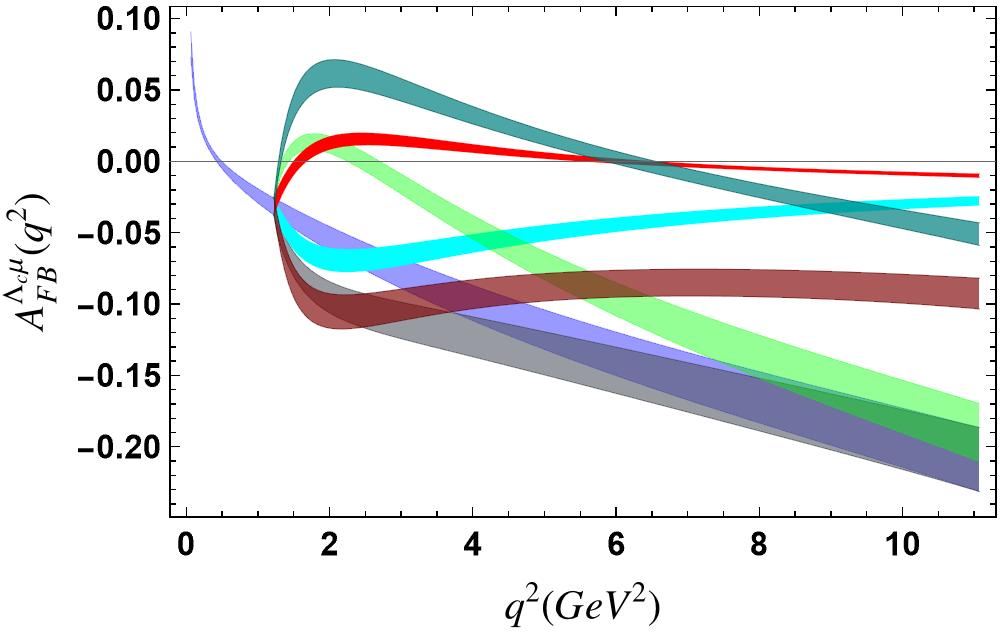}}
	\vspace{-0.1cm}\hspace{0.0002cm}
	\subfloat[]{\includegraphics[scale=0.43]{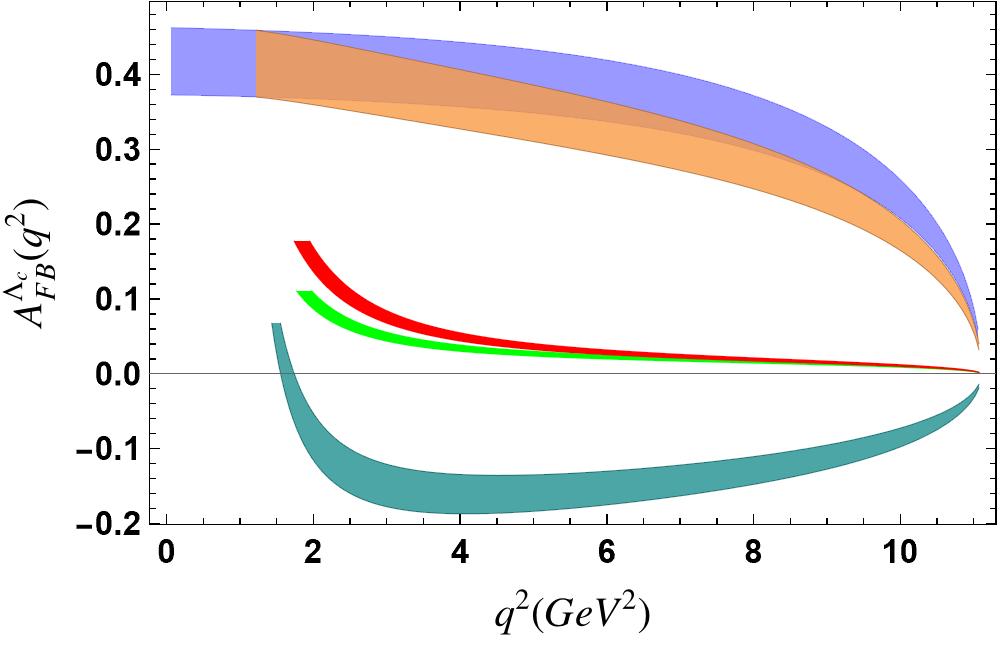}}
	\vspace{-0.1cm}\hspace{0.0002cm}
	\subfloat[]{\includegraphics[scale=0.43]{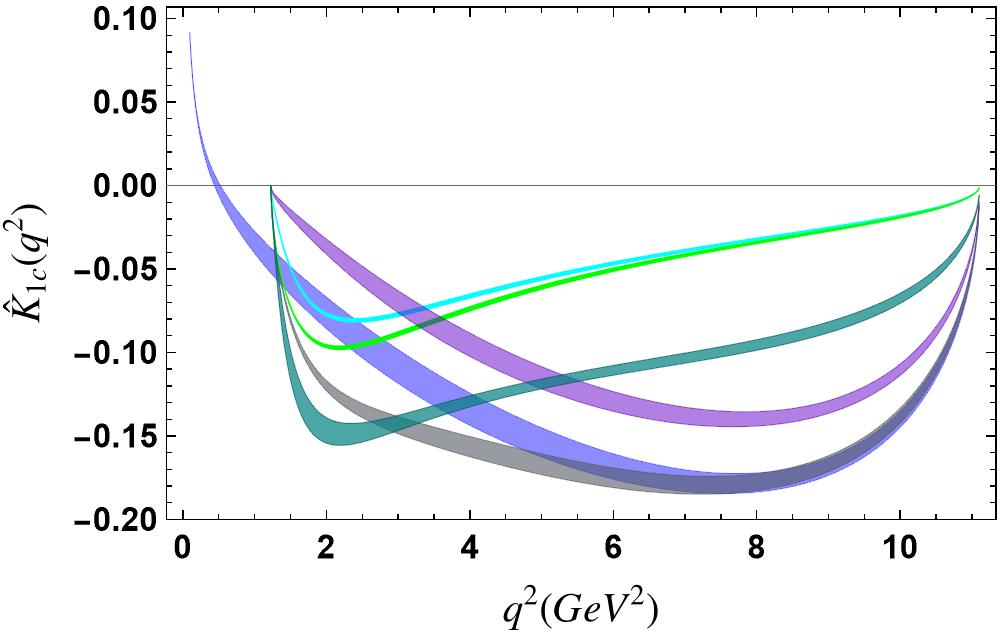}}
	\vspace{-0.1cm}\hspace{0.0003cm}
	\subfloat[]{\includegraphics[scale=0.43]{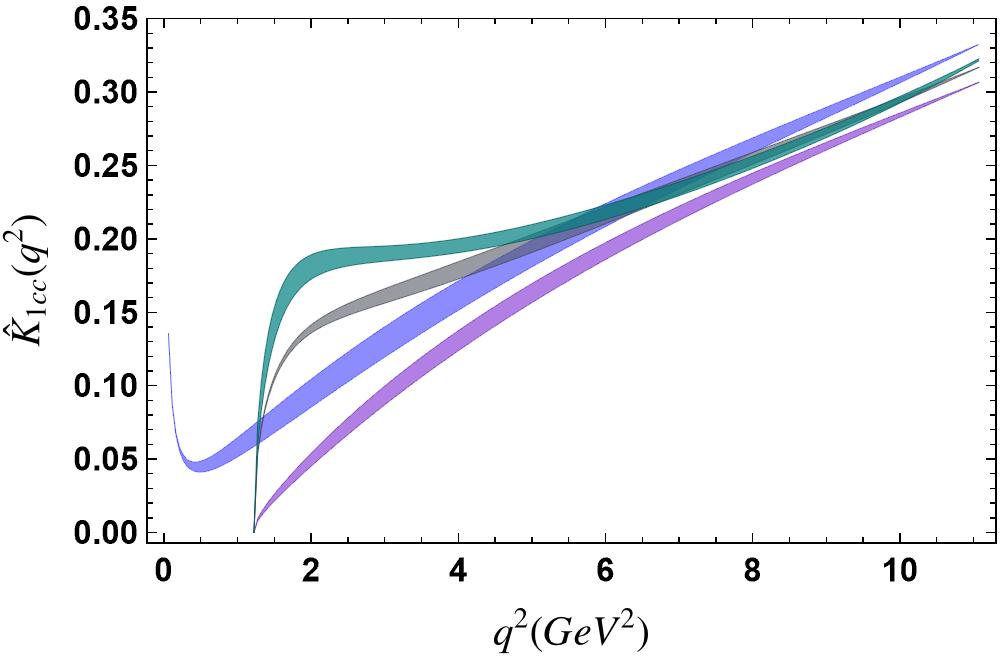}}
	\vspace{-0.1cm}\hspace{0.0002cm}
	\subfloat[]{\includegraphics[scale=0.43]{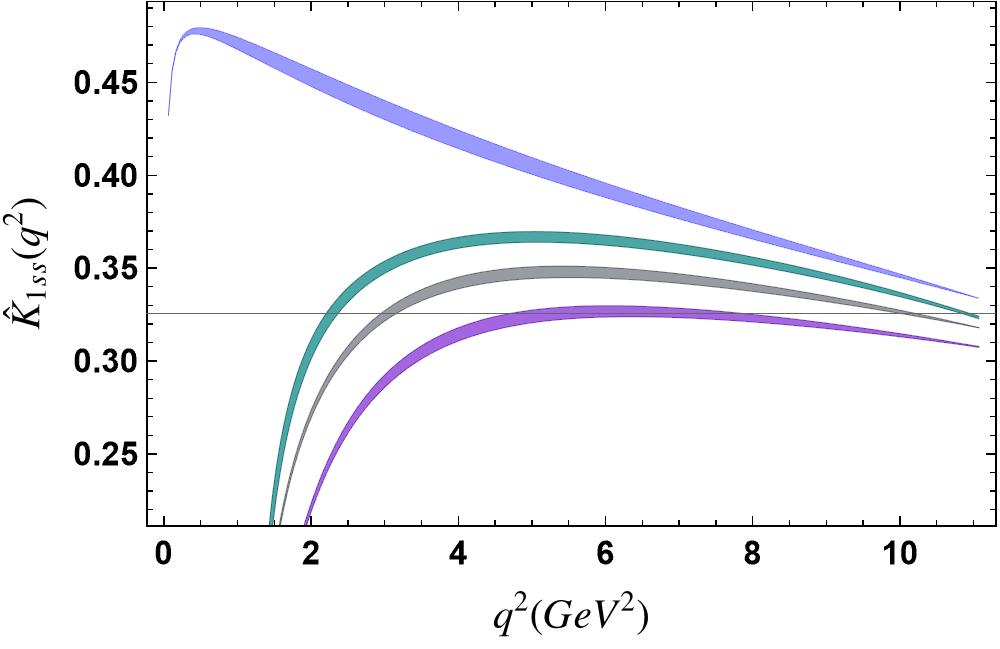}}
	\vspace{-0.1cm}\hspace{0.0002cm}
	\subfloat[]{\includegraphics[scale=0.43]{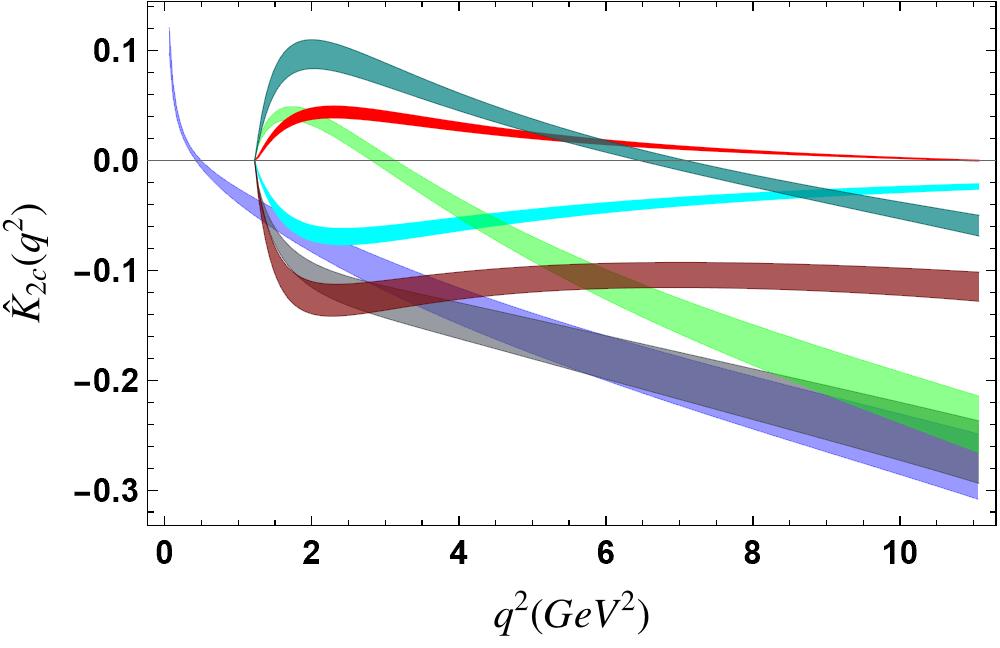}}
	\vspace{-0.1cm}\hspace{0.0002cm} 	
	\subfloat[]{\includegraphics[scale=0.43]{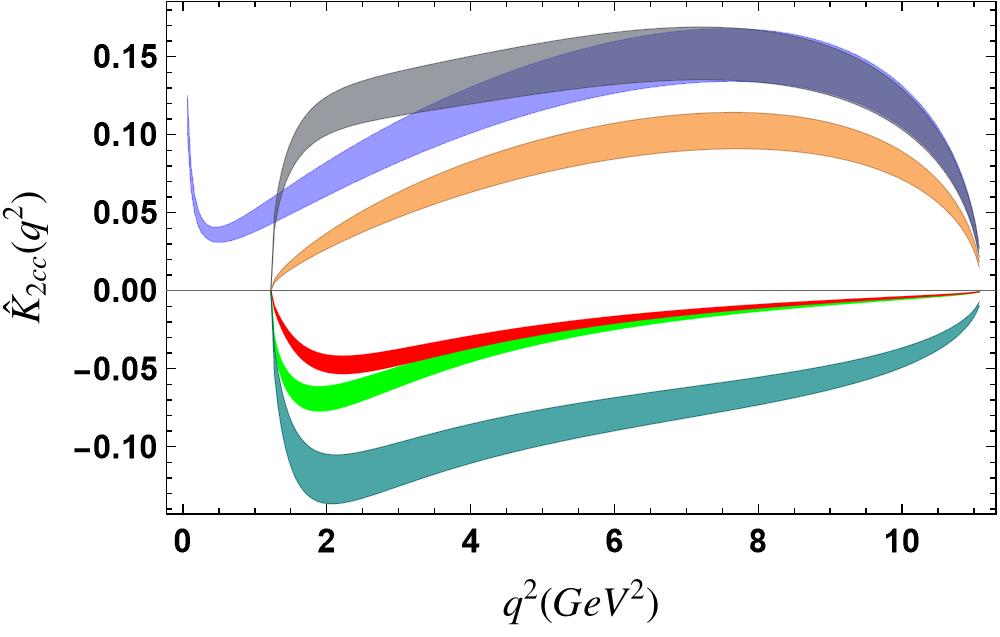}}
	\vspace{-0.1cm}\hspace{0.0002cm}
	\subfloat[]{\includegraphics[scale=0.43]{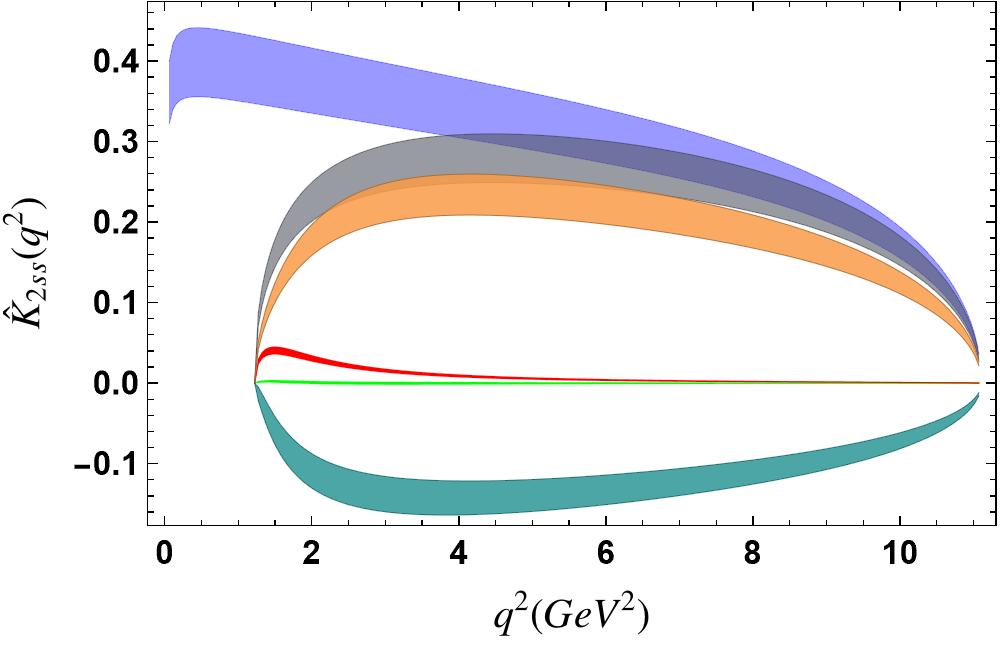}}
	\vspace{-0.1cm}\hspace{0.0002cm}
	\subfloat[]{\includegraphics[scale=0.43]{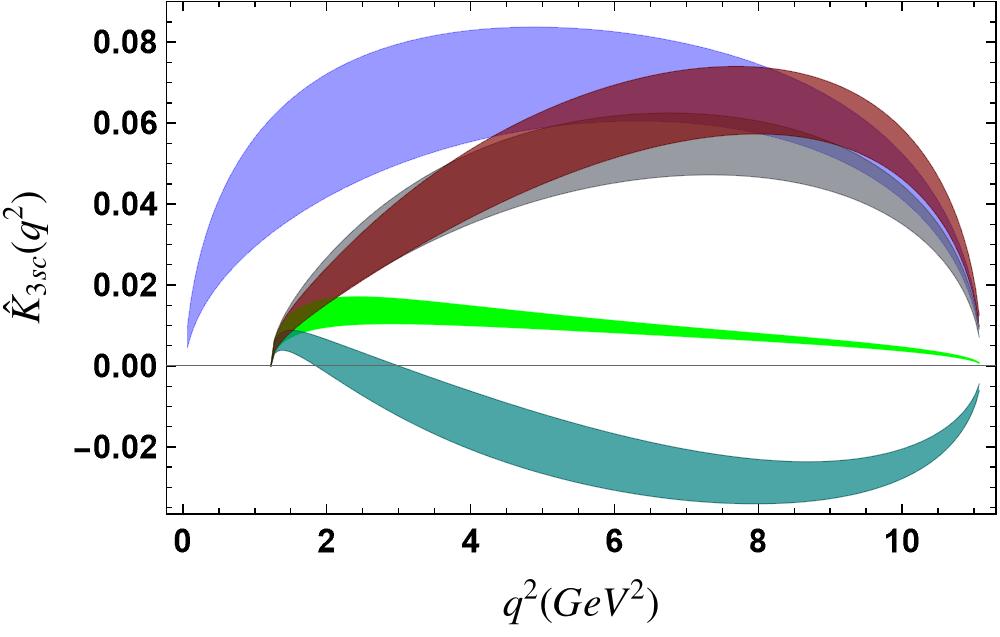}}
	\vspace{-0.1cm}\hspace{0.0002cm}
	\subfloat[]{\includegraphics[scale=0.43]{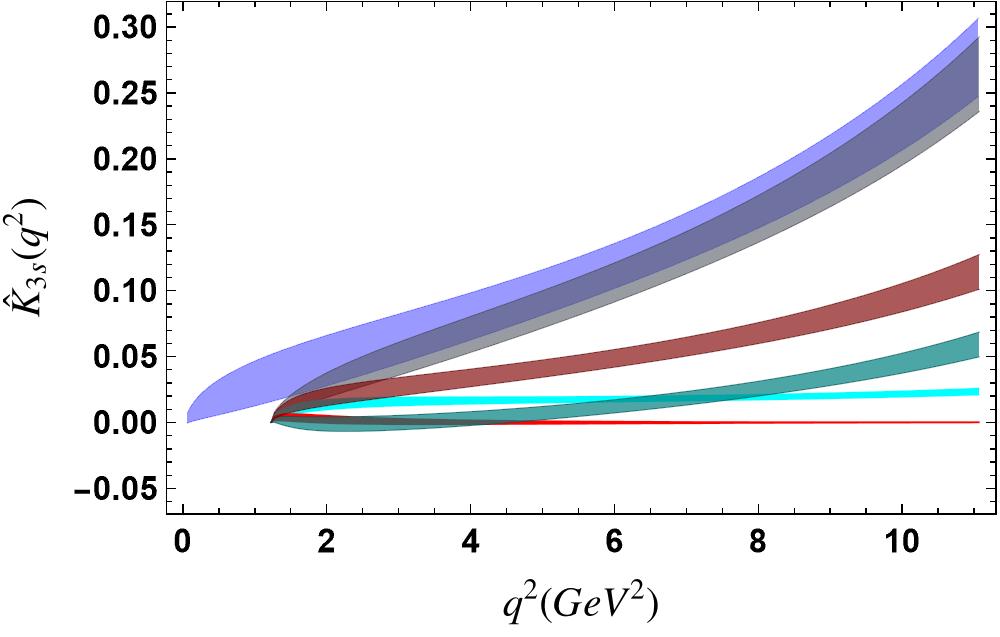}}
	\vspace{0.25cm}
	\hspace{0.0002cm}{\includegraphics[scale=0.7]{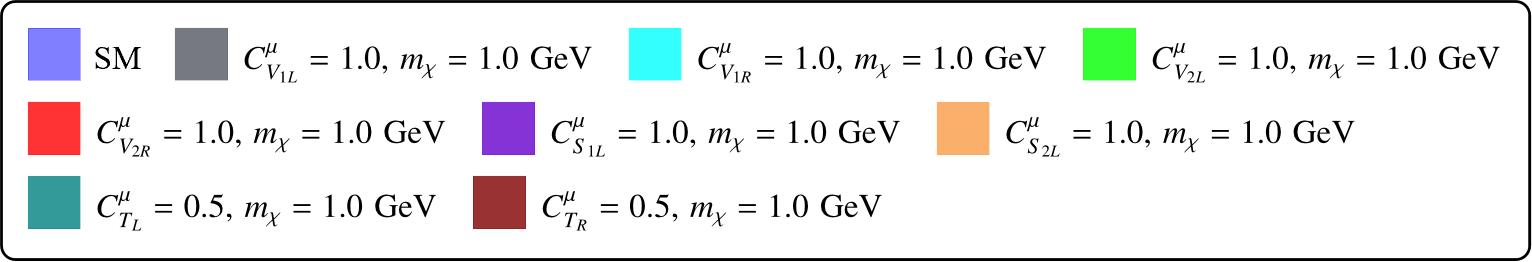}}
	\caption{The $q^2$-distribution of the angular observables in the one-operator scenarios, considering both left- and right-handed currents for benchmark values of WCs and $m_{\chi}$. For non-zero mass of the invisible particle, the region $q^2 \in [m_{\ell}^2, \,  (m_{\ell} + m_{\chi})^2]$ follows purely SM distribution, hence, not shown in the plots. The colored band shows the corresponding $1\sigma$ uncertainties of the observables.}
	\label{fig:Lc_fourfld_plots_mu1pt0}
\end{figure*}

We investigate the imprints of new physics (NP) and the effects of a massive invisible particle, $\chi$, on the angular observables discussed above. Our primary goal is to identify the observables, which are genuinely sensitive to the mass of $\chi$. Not all angular observables show such sensitivity. Also, this dependence is not universal across all NP scenarios. Therefore, in this work, we focus exclusively on the angular observables and NP frameworks where the effects of a non-zero $\chi$ mass are significant. We have a complementary objective: to isolate observables that can discriminate between left- and right-handed lepton currents. These observables are valuable for disentangling the chiral structure of the underlying interactions. They are also important for distinguishing NP scenarios involving dark-sector fermions with both left- and right-chiral couplings from frameworks such as sterile-neutrino or right-handed neutrino models, which typically have a more restricted chiral structure.

In fig.~\ref{fig:Lc_fourfld_plots_mu1pt0}, we present the $q^2$ distributions of the angular observables that are sensitive to $m_{\chi}$ and/or can discriminate between left- and right-handed lepton currents. These distributions are shown for representative benchmark values of $m_{\chi}$, with WCs $\lesssim \mathcal{O}(1)$, consistent with current state-of-the-art constraints~\cite{Kolay:2026mgv, Kolay:2026bjm}. While the plots are largely self-explanatory, we summarise the key observations below:
Although all observables considered here are sensitive to both tensor and vector/axial-vector interactions, only a few can effectively discriminate between these Lorentz structures. In particular, $A_{FB}^{\Lambda_c}(q^2)$, $\hat{K}_{2ss}(q^2)$, and $\hat{K}_{3sc}(q^2)$ exhibit a distinctive behaviour: tensor interactions produce negative shifts and may even render these observables negative, whereas vector and axial-vector interactions can generate sizeable deviations while generally keeping the observables positive or close to zero. In addition, $d\Gamma/dq^2$ shows a much stronger sensitivity to the tensor coupling than to vector and axial-vector operators. Consequently, a correlated study of these observables can provide a robust means of disentangling tensor-current effects from other new-physics contributions.

The tensor couplings $C_{T_L}$ and $C_{T_R}$ affect almost all observables considered in this study. However, $\hat{K}_{2ss}(q^2)$, $\hat{K}_{2cc}(q^2)$, and $\hat{K}_{3sc}(q^2)$ are insensitive to $C_{T_R}$, making them particularly useful for distinguishing left-handed tensor interactions from right-handed ones. In addition, the distinct patterns of NP effects in $A_{FB}^{\Lambda_c\mu}(q^2)$ and $\hat{K}_{2c}(q^2)$ provide further discriminatory power between $C_{T_L}$ and $C_{T_R}$. Consequently, a combined analysis of these observables can serve as an effective probe of the chiral structure of tensor currents.  

All scalar and pseudoscalar interactions significantly affect $\hat{K}_{1c}(q^2)$ and $\hat{K}_{1cc}(q^2)$. However, the different scalar and pseudoscalar scenarios produce qualitatively similar effects in these observables, limiting their ability to distinguish among the various scalar operators. In contrast, the only contributions of $C_{{S_2}_{L,R}}$ can be clearly separated from the corresponding SM predictions in $\hat{K}_{2cc}(q^2)$ and, particularly at low $q^2$, in $\hat{K}_{2ss}(q^2)$. These observables therefore provide sensitivity to the chirality of the quark current, allowing left- and right-handed scalar/pseudoscalar interactions to be distinguished, although they remain insensitive to the chirality of the lepton current. Moreover, the characteristic NP patterns induced by scalar/pseudoscalar interactions in different angular observables are clearly distinguished from those generated by tensor and vector operators.

Once the effects of vector/axial-vector interactions have been distinguished from those of scalar/pseudoscalar and tensor operators, it becomes important to further discriminate among the different vector and axial-vector structures themselves. The differential decay rate $d\Gamma/dq^2$ exhibits sizeable deviations from the SM prediction in the presence of $C_{{V_1}_{L,R}}$, while the effects of $C_{{V_2}_{L,R}}$ remain largely indistinguishable from the SM. In contrast, $A_{FB}^{\Lambda_c}(q^2)$, $\hat{K}_{2ss}(q^2)$, and $\hat{K}_{2cc}(q^2)$ show observable deviations only for $C_{{V_2}_{L,R}}$, thereby providing a clean separation between the $V_1$- and $V_2$-type operators. Although these observables do not distinguish the chirality of the lepton current, additional discrimination is provided by $A_{FB}^{\mu}(q^2)$ and $\hat{K}_{1c}(q^2)$, which can separate the impact of $C_{{V_1}_{L}}$ from $C_{{V_1}_{R}}$ and from $C_{{V_2}_{L}}$. The observables $A_{FB}^{\Lambda_c\mu}(q^2)$ and $\hat{K}_{2c}(q^2)$ can effectively distinguish the effects of $C_{{V_1}_{R}}$ from those of $C_{{V_2}_{L}}$. Furthermore, these observables provide additional discriminatory power by separating the impact of $C_{{V_2}_{R}}$ from those arising from $C_{{V_2}_{L}}$ and $C_{{V_1}_{L}}$. Consequently, a combined study of the $q^2$ distributions of these observables can effectively disentangle all four vector/axial-vector operators and probe the chiral structure of the underlying quark and lepton currents. In appendix~\ref{appsec:angular_Obs2}, we present the $q^2$ distributions of the angular observables sensitive to $m_\chi$ for another representative benchmark point $m_\chi=0.5 \text{GeV}$ with WC $\lesssim \mathcal{O}(1)$. Across all NP scenarios discussed we find qualitatively similar effects, albeit with reduced sensitivity in this case for the angular observables.

\paragraph{\underline{Summary}:}
In this work, we demonstrate that a combined analysis of the differential decay rate and angular observables in semileptonic $\Lambda_b$ decays provides a powerful framework for determining the Lorentz and chiral structure of possible new-physics interactions. We identify observables that can cleanly discriminate tensor, scalar/pseudoscalar, and vector/axial-vector contributions through their characteristic patterns of deviations from the SM predictions. Furthermore, we show that specific combinations of angular observables can distinguish left- and right-chiral quark and lepton currents, enabling a detailed probe of the underlying new-physics structure. A particularly important result is the sensitivity of several observables to the chirality of the lepton current. This offers a unique opportunity to distinguish scenarios involving sterile neutrinos or right-handed leptonic currents from those mediated solely by the standard left-handed weak interaction. Therefore, future precision measurements of the $q^2$ distributions and angular observables in $\Lambda_b$ decays can not only reveal the presence of new physics but also determine its Lorentz structure and chiral origin.

\acknowledgments
S.S. acknowledges financial support under ANRF Grant No. ANRF/IRG/2024/000256/PS.
\appendix
\onecolumngrid
\section{Decay Amplitude}~\label{appsec:helicityamp}
Assuming a factorization between the hadronic and leptonic components, we can express the three-body helicity amplitude $\mathcal{M}^{\lambda_\ell}_{\lambda_{\Lambda_c}}(\Lb \to \Lc \ell \bar{\chi})$ as:
{\small
\begin{align}\label{eq:amplitude_gen}
	\mathcal{M}^{\lambda_\ell}_{\lambda_{\Lambda_c}}(\Lb \to \Lc \ell \bar{\chi})=& \frac{G_F V_{cb}}{\sqrt{2}} 
	\sum_{\alpha, \, \beta=L,R}\biggl[
	H^{SP-\alpha}_{\lambda_{\Lc}} \, L^{SP-\beta, \, \lambda_{\ell}}_{\lambda_{\chi}} + \sum_{\lambda} \eta_{\lambda} \, H^{VA-\alpha}_{\lambda_{\Lc}, \, \lambda} \, L^{VA-\beta, \, \lambda_{\ell}}_{\lambda_{\chi}, \, \lambda} + \,  \sum_{\lambda}\sum_{\lambda^{\prime}} \delta_{\alpha \, \beta} \, \eta_{\lambda} \eta_{\lambda^{\prime}} \, \, H^{T-\alpha}_{\lambda_{\Lc}, \, \lambda, \, \lambda^{\prime}} \, L^{T-\beta, \, \lambda_{\ell}}_{\lambda_{\chi}, \, \lambda, \, \lambda^{\prime}}\biggr].
\end{align}
}
In the above equation, $H$ and $L$ correspond to the hadronic and leptonic amplitudes, respectively. Here, $ \lambda,\, \lambda' = (t,+,-,0) $,  are the helicity of the virtual gauge boson, while  $\eta_{\lambda}$ being the metric factor. Tensor operator with different lepton and quark chiralities vanishes identically. This is direct consequence of the Dirac-algebra identity $\gamma_5 \sigma^{\mu \nu}=\frac{i}{2}\epsilon^{\mu \nu \alpha\beta} \sigma_{\alpha \beta}$, which implies $\sigma_{\mu \nu} \otimes  \sigma^{\mu \nu} \gamma_5=\sigma_{\mu \nu} \gamma_5 \otimes \sigma^{\mu \nu}$ and $\sigma_{\mu \nu} \gamma_5  \otimes  \sigma^{\mu \nu} \gamma_5=\sigma_{\mu \nu} \otimes \sigma^{\mu \nu}$. Here, ($\lambda$, $\lambda^\prime$) indicate the helicity of the virtual vector boson, $\lambda_{\Lambda_c}$ and $\lambda_\ell$ are the helicities of the  $\Lambda_c$ baryon and $\ell$ lepton, respectively, and $\eta_\lambda=1$ for $\lambda=t$ and $\eta_\lambda=-1$ for $\lambda=0,\pm 1$. The semi-leptonic decay $\Lambda_b \to \Lambda_{c} \ell^- \bar{X}_{\ell}$ can be considered as two subsequent 2-body decays, such as the decay $\Lambda_b \to \Lambda_{c} W^*$, followed by a subsequent decay of the off-shell $W^*$ to $\ell^- \bar{X}_{\ell}$. The off-shell $W^*$ has four helicities, with two angular momentum $J=0,1$ in the rest frame of $W^*$, namely $\lambda = \pm 1,0$ ($J=1$) and $\lambda =0$ ($J=0$), only the off-shell $W^*$ has time-like component. To distinguish the two $\lambda=0$ states, we have adopted the notation $\lambda=0$ for $J=1$ and $\lambda=t$ for $J=0$. The scalar-type, vector/axial-vector-type, and tensor-type hadronic helicity amplitudes are defined as
\begin{eqnarray}~\label{eq:helicity_def1}
	H^{SP-L}_{\lambda_{\Lambda_c},\lambda=0} &&= H^{SL}_{\lambda_{\Lambda_c}, \lambda=0 }+ H^{PL}_{\lambda_{\Lambda_c},\lambda=0} \, , \qquad 
	H^{SP-R}_{\lambda_{\Lambda_c},\lambda=0} = H^{SR}_{\lambda_{\Lambda_c}, \lambda=0} + H^{PR}_{\lambda_{\Lambda_c},\lambda=0} \, , \nn\\ 
	 H^{VA-L}_{\lambda_{\Lambda_c},\lambda} && =  H^{VL}_{\lambda_{\Lambda_c}, \lambda} - H^{AL}_{\lambda_{\Lambda_c},\lambda} \, , \qquad \qquad
	H^{VA-R}_{\lambda_{\Lambda_c},\lambda}=  H^{VR}_{\lambda_{\Lambda_c}, \lambda} - H^{AR}_{\lambda_{\Lambda_c},\lambda} \, ,\nn\\
	H^{T-L,{\lambda_{\lb}}}_{\lambda_{\lc}, \lambda ,\lambda^{\prime}} &&= H^{TL1,{\lambda_{\lb}}}_{\lambda_{\lc}, \lambda ,\lambda^{\prime}} - H^{TL2,{\lambda_{\lb}}}_{\lambda_{\lc},\lambda ,\lambda^{\prime}} \, , \qquad	H^{T-R,{\lambda_{\lb}}}_{\lambda_{\lc}, \lambda ,\lambda^{\prime}}= H^{TR1,{\lambda_{\lb}}}_{\lambda_{\lc},\lambda ,\lambda^{\prime}}+ H^{TR2,{\lambda_{\lb}}}_{\lambda_{\lc},\lambda ,\lambda^{\prime}} \, .
\end{eqnarray}
Where,
\begin{center}
\begin{eqnarray}~\label{eq:helicity_def2}
	H^{SL}_{\lambda_{\Lambda_c},\lambda=0} & = & (C_{S_{1L}}+C_{S_{2L}}) \bra{\lc}\bar{c} b\ket{\lb} \, , \qquad \qquad \qquad \, \,
	H^{PL}_{\lambda_{\Lambda_c},\lambda=0}=(C_{S_{1L}}-C_{S_{2L}})\bra{\lc}\bar{c}\gamma_5 b\ket{\lb} \, , \nn\\
	H^{SR}_{\lambda_{\Lambda_c},\lambda=0} &=& (C_{S_{1R}}+C_{S_{2R}}) \bra{\lc}\bar{c} b\ket{\lb} \, ,\qquad \qquad \qquad \, \,
	H^{PR}_{\lambda_{\Lambda_c},\lambda=0}=(C_{S_{1R}}-C_{S_{2R}})\bra{\lc} \bar{c}\gamma_5 b\ket{\lb} \, , \nn \\
	H^{VL}_{\lambda_{\Lambda_c},\lambda} & = &(C_{V_{1L}}+C_{V_{2L}})\,\epsilon^*_{\mu}(\lambda)\bra{\lc}\bar{c}\gamma^{\mu} b\ket{\lb} \, , \qquad \qquad
	H^{AL}_{\lambda_{\Lambda_c},\lambda}= (C_{V_{1L}}-C_{V_{2L}})\,\epsilon^*_{\mu}(\lambda)\bra{\lc}\bar{c}\gamma^{\mu}\gamma_5 b\ket{\lb} \, , \nn\\
	H^{VR}_{\lambda_{\Lambda_c},\lambda}&=&(C_{V_{1R}}+C_{V_{2R}})\,\epsilon^*_{\mu}(\lambda)\bra{\lc}\bar{c}\gamma^{\mu} b\ket{\lb} \, , \qquad \qquad
	H^{AR}_{\lambda_{\Lambda_c},\lambda}= (C_{V_{1R}}-C_{V_{2R}})\,\epsilon^*_{\mu}(\lambda)\bra{\lc}\bar{c}\gamma^{\mu}\gamma_5 b\ket{\lb} \, , \nn \\
	H^{TL1,{\lambda_{\lb}}}_{\lambda_{\lc},\lambda,\lambda^{\prime}}&=&
	C_{T_L}\:\epsilon^*_{\mu}(\lambda)\epsilon^*_{\nu}(\lambda^{\prime}) \bra{\lc}\bar{c}i\sigma^{\mu \nu} b\ket{\lb} \, , \qquad \qquad
	H^{TL2,{\lambda_{\lb}}}_{\lambda_{\lc},\lambda ,\lambda^{\prime}}= C_{T_L}\:\epsilon^*_{\mu}(\lambda)\epsilon^*_{\nu} (\lambda^{\prime})\bra{\lc}\bar{c}i\sigma^{\mu \nu}\gamma_5 b\ket{\lb} \, , \nn\\
	H^{TR1,{\lambda_{\lb}}}_{\lambda_{\lc},\lambda ,\lambda^{\prime}}& = &
	C_{T_R}\:\epsilon^*_{\mu}(\lambda)\epsilon^*_{\nu}(\lambda^{\prime}) \bra{\lc}\bar{c}i\sigma^{\mu \nu} b\ket{\lb} \, , \qquad \qquad
	H^{TR2,{\lambda_{\lb}}}_{\lambda_{\lc},\lambda ,\lambda^{\prime}}= C_{T_R}\:\epsilon^*_{\mu}(\lambda)\epsilon^*_{\nu}(\lambda^{\prime}) \bra{\lc}\bar{c}i\sigma^{\mu \nu}\gamma_5 b\ket{\lb} \, .
\end{eqnarray}
\end{center}
Similarly, we also define the leptonic helicity amplitudes for charged lepton and the dark-fermion $\chi$ as follows: 	
\begin{align}\label{eq:leptonic_helicity_amp}
	L^{\lambda_\ell, \, SP-L(R)}_{\lambda_{\chi_\ell}} (q^2,\cos \theta_{\ell})=&\bra{\ell\bar{\chi}_\ell}\bar{\ell} (1 \mp \gamma_5)\chi_\ell\ket{0} = \bar{u}_\ell (1 \mp \gamma_5) v_{\bar{\chi}_\ell} \, ,\nn \\
    L^{\lambda_\ell, \, VA-L(R)}_{\lambda_{\chi},\lambda} (q^2,\cos \theta_{\ell}) = & \epsilon_\mu (\lambda)\bra{\ell\bar{\chi}_\ell}\bar{\ell}\gamma^\mu (1 \mp \gamma_5) \chi_\ell\ket{0}=\epsilon_\mu (\lambda) \bar{u}_\ell \gamma^\mu(1 \mp \gamma_5) v_{\bar{\chi}_\ell}  \, ,\nn \\
	L^{\lambda_\ell, \, T-L(R)}_{ \lambda_{\chi_\ell}, \, \lambda ,\lambda^{\prime}} (q^2,\cos \theta_{\ell}) =& -i\epsilon_\mu (\lambda)\epsilon_\nu (\lambda^\prime)\bra{\ell\bar{\chi}_\ell}\bar{\ell}\sigma^{\mu \nu} (1\mp \gamma_5)\chi_\ell\ket{0}=-i\epsilon_\mu (\lambda)\epsilon_\nu (\lambda^\prime)\bar{u}_\ell \sigma^{\mu \nu} (1\mp \gamma_5) v_{\bar{\chi}_\ell} \, ,
\end{align}
where, $\epsilon^{\mu}(\lambda)$ are the polarization vectors of the virtual vector boson. The explicit expressions for the hadronic and leptonic helicity amplitudes are derived and presented below.
\subsection{Transition Matrix Elements and Form Factors}
The transition matrix elements of the vector and axial vector current are as follows
\begin{eqnarray}
	\bra{\lc}\bar{c}\gamma^\mu b\ket{\lb}&=&\bar{u}_{\lc}\Big[ f_0 (q^2)(m_{\lb} - m_{\lc})\frac{q^\mu}{q^2} 
	+f_+ (q^2)\frac{m_{\lb} + m_{\lc}}{Q_+}(p_{\lb}^{\mu} +p_{\lc}^{\mu}-(m_{\lb} ^2 - m_{\lc} ^2)\frac{q^\mu}{q^2}) \nonumber\\
	&&+f_\perp (q^2)(\gamma^\mu - \frac{2m_{\lc}}{Q_+}p_{\lb}^{\mu} - \frac{2m_{\lb}}{Q_+}p_{\lc}^{\mu})\Big]u_{\lb}, \label{eq:VFF} \\
	\bra{\lc}\bar{c}\gamma^\mu \gamma_5 b\ket{\lb}&=&-\bar{u}_{\lc}\gamma_5\Big[ g_0 (q^2)(m_{\lb} + m_{\lc})\frac{q^\mu}{q^2} \nonumber\\
	&&+g_+ (q^2)\frac{m_{\lb} - m_{\lc}}{Q_-}(p_{\lb}^{\mu} +p_{\lc}^{\mu}-(m_{\lb} ^2 - m_{\lc} ^2)\frac{q^\mu}{q^2})\nonumber\\
	&&+g_\perp (q^2)(\gamma^\mu + \frac{2m_{\lc}}{Q_-}p_{\lb}^{\mu} - \frac{2m_{\lb}}{Q_-}p_{\lc}^{\mu})\Big]u_{\lb} \label{eq:AFF}.
\end{eqnarray}

The matrix elements of the scalar and pseudo-scalar currents can be obtained from the vector and axial vector matrix elements using the equations of motion:
\begin{align}
	\nonumber \bra{\lc}\bar{c} b\ket{\lb} =& \frac{q_\mu}{m_b-m_c}\bra{\lc}\bar{c}\gamma^\mu b\ket{\lb} \\
	=& f_0(q^2)  \frac{m_{\lb} - m_{\lc}}{m_b-m_c} \bar{u}_{\lc}u_{\lb}, \\
	\nonumber \bra{\lc}\bar{c}\gamma_5 b\ket{\lb} =& \frac{q_\mu}{m_b+m_c}\bra{\lc}\bar{c}\gamma^\mu\gamma_5 b\ket{\lb} \\
	=& g_0(q^2)  \frac{m_{\lb} + m_{\lc}}{m_b+m_c} \bar{u}_{\lc}\gamma_5 u_{\lb}.
\end{align}
The matrix element of the tensor current can be written in terms of four form factors $h_+$, $h_\perp$, $\widetilde{h}_+$, $\widetilde{h}_\perp$ as
\begin{align}
	\bra{\lc}\bar{c}i\sigma^{\mu\nu} b\ket{\lb}=&\bar{u}_{\lc}\Big[2h_+(q^2)\frac{p_{\lb}^\mu p_{\lc}^{ \nu}-p_{\lb}^\nu p_{\lc}^{\mu}}{Q_+} \nonumber\\
	&+h_\perp (q^2)\Big(\frac{m_{\lb}+m_{\lc}}{q^2}(q^\mu \gamma^\nu -q^\nu \gamma^\mu)-2(\frac{1}{q^2}+\frac{1}{Q_+})(p_{\lb}^\mu p_{\lc}^{\nu}-p_{\lb}^\nu p_{\lc}^{\mu}) \Big) \nonumber\\
	&+\widetilde{h}_+ (q^2)\Big(i\sigma^{\mu \nu}-\frac{2}{Q_-}(m_{\lb}(p_{\lc}^{\mu}\gamma^\nu -p_{\lc}^{\nu}\gamma^\mu)\nonumber\\
	&-m_{\lc}(p_{\lb}^\mu \gamma^\nu -p_{\lb}^\nu \gamma^\mu)+p_{\lb}^\mu p_{\lc}^{\nu}-p_{\lb}^\nu p_{\lc}^{\mu}) \Big) \nonumber\\
	&+\widetilde{h}_\perp(q^2) \frac{m_{\lb}-m_{\lc}}{q^2 Q_-}\Big((m_{\lb}^2-m_{\lc}^2-q^2)(\gamma^\mu p_{\lb}^\nu - \gamma^\nu p_{\lb}^\mu)\nonumber\\
	&-(m_{\lb}^2-m_{\lc}^2+q^2)(\gamma^\mu p_{\lc}^{\nu}-\gamma^\nu p_{\lc}^{\mu})+2(m_{\lb}-m_{\lc})(p_{\lb}^\mu p_{\lc}^{\nu}-p_{\lb}^\nu p_{\lc}^{\mu}) \Big)
	\Big]u_{\lb}. \nonumber \\ \label{eq:TFF}
\end{align}
The matrix elements of the pseudo-tensor current $\bar{c}i\sigma^{\mu\nu}\gamma_5 b$ can be obtained from the above equation by using the identity
\begin{align}
	\sigma^{\mu \nu}\gamma_{5}=\frac{i}{2}\epsilon^{\mu \nu \alpha \beta}\sigma_{\alpha \beta}.
\end{align}
Here, we adopt $\epsilon^{0123}=+1$ for evaluating the pseudo-tensor matrix elements in this work.
\subsection{Kinematics in $\Lambda_b$ rest frame}~\label{appendix:Kin_Lbframe}
To calculate the hadronic helicity amplitudes, we work in the $\Lambda_b$ rest frame, positioning the three-momentum of the $\Lambda_c$ along the $+z$ direction and the three-momentum of the virtual vector boson along the $-z$ direction. The baryon spinors are then expressed as follows:
\begin{eqnarray}
	u_{\lb}\Big(\pm \frac{1}{2}, p_{\lb} \Big) &=& \sqrt{2m_{\lb}} 
	\left(
	\begin{array}{l}
		\chi_\pm \\
		~0 \\
	\end{array}
	\right) \,, \nonumber\\
	\bar u_{\Lambda_c}\Big( \pm \frac{1}{2}, p_{\Lambda_c}\Big) &=& \sqrt{E_{\Lambda_c} + m_{\Lambda_c}}
	\Big( \chi_\pm^\dagger, \frac{\mp |{\rm p_{\Lambda_c}}|}{E_{\Lambda_c} + m_{\Lambda_c}}  
	\chi_\pm^\dagger \Big)
\end{eqnarray}
where $\chi_+ = \left(
\begin{array}{l}
	1 \\
	0 \\
\end{array} \right)$ 
and $\chi_- = \left(
\begin{array}{l}
	0 \\
	1 \\
\end{array} \right)$ are two-component Pauli spinors.    
In the $\Lambda_b$ rest frame, the four-momenta of $\Lambda_b(p_{\lb})$, $\Lambda_c(p_{\Lambda_c})$ and $W^*(q)$ are given as
\begin{eqnarray}
	p_{\lb}^{\mu} &=& \left(m_{\lb}, 0, 0, 0\right)\,, \nonumber \\
	p_{\Lambda_c}^{\mu} &=& \left(E_{\Lambda_c}, 0, 0, |\rm q|\right) \,,\nonumber \\
	q^{\mu} &=& \left(q_0, 0, 0, -|\rm q|\right)\,.
\end{eqnarray}
The polarization vectors of virtual vector boson($\epsilon$) are defined in the $\Lambda_b$ rest frame as follows:
\begin{eqnarray}
	\label{}
	\epsilon^{\mu *}(t) = \frac{1}{\sqrt{q^2}}\left(q_0,0,0,-|\rm q|\right) ,\qquad \qquad
	\epsilon^{\mu *}(0) = \frac{1}{\sqrt{q^2}} \left(|\rm q|, 0, 0, -q_0\right) , \qquad \qquad
	\epsilon^{\mu *}(\pm 1) = \frac{1}{\sqrt{2}}\left(0, \pm 1, i, 0\right)
\end{eqnarray}
where,
\begin{equation}
	|\rm p_{\Lambda_c}|=|\rm q|=\frac{\sqrt{Q_+ Q_-}}{2 m_{\lb}},\qquad
	E_{\Lambda_c}=\frac{m_{\lb}^2+m_{\Lambda_c}^2-q^2}{2m_{\lb}},\qquad 
	q_0= \frac{m_{\lb}^2-m_{\Lambda_c}^2+q^2}{2m_{\lb}}. 
\end{equation} 
The vectors satisfy the following orthonormality and completeness relations
\begin{align}
	&\epsilon^{*\mu}(n) \epsilon_{\mu}(n^{\prime})=g_{n n^{\prime}}, \, \, \sum_{n, n^{\prime}}\epsilon^{*\mu}(n) \epsilon^{\nu}(n^{\prime}) g_{n n^{\prime}}=g^{\mu \nu}, \, \,~~\text{with}~~~~ \, n,n^{\prime}=t, \pm ,0 \nn \\
	&\epsilon^{\mu}(\lambda) \cdot q_{\mu}=0  \, \,~~\text{with}~~~~ \, \lambda= \pm ,0.
\end{align}
where, $g_{n n^{\prime}}=\text{diag}(+1,-1,-1,-1)$ and our choice of the metric tensor is $g^{\mu \nu}=\text{diag}(+1,-1,-1,-1)$.
\subsection{Hadronic helicity amplitudes}
Using eqs.~\ref{eq:helicity_def1} and \ref{eq:helicity_def2}, we calculated the hadronic helicity amplitudes with the spinors for $\Lb$ and $\Lc$ in the rest frame of $\Lb$ and the polarization vectors for the virtual vector boson $W^*$. In the following, only the non-vanishing helicity amplitudes are presented.
\subsubsection*{Scalar and pseudo-scalar helicity amplitudes}
 The scalar and pseudo-scalar helicity amplitudes associated with the form factors $f_i(q^2), g_i(q^2)$ are
\begin{align}
	H^{SP-L}_{1/2,0} &= f_0(q^2)(C_{S_{1L}}+C_{S_{2L}}) \frac{\sqrt{Q_+}}{m_b-m_c}(m_{\Lambda_b}-m_{\Lambda_c}) - g_0(q^2)(C_{S_{1L}}-C_{S_{2L}})\frac{\sqrt{Q_-}}{m_b+m_c}(m_{\Lambda_b}+m_{\Lambda_c}), \nn \\[2pt]
	H^{SP-L}_{-1/2,0} &= f_0(q^2)(C_{S_{1L}}+C_{S_{2L}})\frac{\sqrt{Q_+}}{m_b-m_c}(m_{\Lambda_b}-m_{\Lambda_c}) + g_0(q^2)(C_{S_{1L}}-C_{S_{2L}})\frac{\sqrt{Q_-}}{m_b+m_c}(m_{\Lambda_b}+m_{\Lambda_c}), \nn \\[6pt]
	H^{SP-R}_{1/2,0} &= f_0(q^2)(C_{S_{1R}}+C_{S_{2R}}) \frac{\sqrt{Q_+}}{m_b-m_c}(m_{\Lambda_b}-m_{\Lambda_c})  - g_0(q^2)(C_{S_{1R}}-C_{S_{2R}})\frac{\sqrt{Q_-}}{m_b+m_c}(m_{\Lambda_b}+m_{\Lambda_c}), \nn \\[6pt]
	H^{SP-R}_{-1/2,0} &= f_0(q^2)(C_{S_{1R}}+C_{S_{2R}})\frac{\sqrt{Q_+}}{m_b-m_c}(m_{\Lambda_b}-m_{\Lambda_c}) + g_0(q^2)(C_{S_{1R}}-C_{S_{2R}})\frac{\sqrt{Q_-}}{m_b+m_c}(m_{\Lambda_b}+m_{\Lambda_c}).
\end{align}
\subsubsection*{Vector and axial-vector helicity amplitudes}
For the vector and axial-vector helicity amplitudes, we find
	\begin{eqnarray}
		H^{VA-L}_{1/2,0}=&& f_+(q^2) (C_{V_{1L}}+C_{V_{2L}})\frac{\sqrt{Q_-}}{\sqrt{q^2}}(m_{\lb}+m_{\lc})-g_+(q^2) (C_{V_{1L}}-C_{V_{2L}})\frac{\sqrt{Q_+}}{\sqrt{q^2}}(m_{\lb}-m_{\lc})
	     , \nn \\[4 pt]
		H^{VA-L}_{1/2,+1}= && -f_\perp(q^2) (C_{V_{1L}}+C_{V_{2L}})\sqrt{2Q_-} +g_\perp(q^2) (C_{V_{1L}}-C_{V_{2L}})\sqrt{2Q_+}, \nn \\[4pt]
		H^{VA-L}_{1/2,t}= && f_0(q^2)(C_{V_{1L}}+C_{V_{2L}})\frac{\sqrt{Q_+}}{\sqrt{q^2}}(m_{\lb}-m_{\lc})-g_0(q^2)(C_{V_{1L}}-C_{V_{2L}})\frac{\sqrt{Q_-}}{\sqrt{q^2}}(m_{\lb}+m_{\lc}), \nn \\[4pt]
		H^{VA-L}_{-1/2,0}= && f_+(q^2) (C_{V_{1L}}+C_{V_{2L}}) \frac{\sqrt{Q_-}}{\sqrt{q^2}}(m_{\lb}+m_{\lc})+g_+(q^2) (C_{V_{1L}}-C_{V_{2L}})\frac{\sqrt{Q_+}}{\sqrt{q^2}}(m_{\lb}-m_{\lc}), \nn \\[10pt]
		H^{VA-L}_{-1/2,-1}= && -f_\perp(q^2) (C_{V_{1L}}+C_{V_{2L}})\sqrt{2Q_-} -g_\perp(q^2) (C_{V_{1L}}-C_{V_{2L}})\sqrt{2Q_+}, \nn \\[10pt]
		H^{VA-L}_{-1/2,t}= && f_0(q^2) (C_{V_{1L}} +C_{V_{2L}})\frac{\sqrt{Q_+}}{\sqrt{q^2}}(m_{\lb}-m_{\lc}) +g_0(q^2)(C_{V_{1L}} -C_{V_{2L}})\frac{\sqrt{Q_-}}{\sqrt{q^2}}(m_{\lb}+m_{\lc}).
	\end{eqnarray}
and,
	\begin{eqnarray}
	H^{VA-R}_{1/2,0}= && f_+(q^2) (C_{V_{1R}}+C_{V_{2R}})\frac{\sqrt{Q_-}}{\sqrt{q^2}}(m_{\lb}+m_{\lc})-g_+(q^2) (C_{V_{1R}}-C_{V_{2R}})\frac{\sqrt{Q_+}}{\sqrt{q^2}}(m_{\lb}-m_{\lc}), \nn \\[4pt]
	H^{VA-R}_{1/2,+1}= &&-f_\perp(q^2) (C_{V_{1R}}+C_{V_{2R}})\sqrt{2Q_-} +g_\perp(q^2) (C_{V_{1R}}-C_{V_{2R}})\sqrt{2Q_+}, \nn\\[10pt]		
	H^{VA-R}_{1/2,t}= &&f_0(q^2)(C_{V_{1R}}+C_{V_{2R}})\frac{\sqrt{Q_+}}{\sqrt{q^2}}(m_{\lb}-m_{\lc})-g_0(q^2)(C_{V_{1R}}-C_{V_{2R}})\frac{\sqrt{Q_-}}{\sqrt{q^2}}(m_{\lb}+m_{\lc}), \nn \\[4pt]
	H^{VA-R}_{-1/2,0}= &&f_+(q^2) (C_{V_{1R}}+C_{V_{2R}}) \frac{\sqrt{Q_-}}{\sqrt{q^2}}(m_{\lb}+m_{\lc})+g_+(q^2) (C_{V_{1R}}-C_{V_{2R}})\frac{\sqrt{Q_+}}{\sqrt{q^2}}(m_{\lb}-m_{\lc}), \nn \\[4pt]
	H^{VA-R}_{-1/2,-1}= &&-f_\perp(q^2) (C_{V_{1R}}+C_{V_{2R}})\sqrt{2Q_-} -g_\perp(q^2) (C_{V_{1R}}-C_{V_{2R}})\sqrt{2Q_+}, \nn \\[4pt]
	H^{VA-R}_{-1/2,t}= && f_0(q^2) (C_{V_{1R}}+C_{V_{2R}})\frac{\sqrt{Q_+}}{\sqrt{q^2}}(m_{\lb}-m_{\lc}) +g_0(q^2)(C_{V_{1R}}-C_{V_{2R}})\frac{\sqrt{Q_-}}{\sqrt{q^2}}(m_{\lb}+m_{\lc}). 
\end{eqnarray}
We also have the relations $H_{\lambda_{\Lambda_{c}},\lambda}^{VA-L(R)}=H_{\lambda_{\Lambda_{c}},\lambda}^{VL(R)}-H_{\lambda_{\Lambda_{c}},\lambda}^{AL(R)}$ where
\begin{eqnarray}
	H_{\lambda_{\Lambda_{c}},\lambda}^{VL(R)}&=&H_{-\lambda_{\Lambda_{c}},-\lambda}^{VL(R)},\nn \\[10pt]
	H_{\lambda_{\Lambda_{c}},\lambda}^{AL(R)}&=&-H_{-\lambda_{\Lambda_{c}},-\lambda}^{AL(R)}.
\end{eqnarray}
\subsubsection*{Tensor helicity amplitudes}
The non-vanishing tensor helicity amplitudes are
\begin{align}
	H^{T-L,-1/2}_{-1/2,t,0}=& \, \, -C_{T_L}\Big[-h_+(q^2)\sqrt{Q_-}+\widetilde{h}_+(q^2)\sqrt{Q_+}\Big], \nn \\[4pt]
	H^{T-L,+1/2}_{+1/2,t,0}=& \, \, C_{T_L}\Big[h_+(q^2)\sqrt{Q_-}+\widetilde{h}_+(q^2)\sqrt{Q_+}\Big], \nn \\[4pt]
	H^{T-L,-1/2}_{+1/2,t,+1}= & \, \,  -C_{T_L}\frac{\sqrt{2}}{\sqrt{q^2}}\Big[h_\perp(q^2)(m_{\lb}+m_{\lc})\sqrt{Q_-}+\widetilde{h}_\perp(q^2)(m_{\lb}-m_{\lc})\sqrt{Q_+}\Big] , \nn \\[4pt]
	H^{T-L,+1/2}_{-1/2,t,-1}=& \, \, -C_{T_L}\frac{\sqrt{2}}{\sqrt{q^2}}\Big[h_\perp(q^2)(m_{\lb}+m_{\lc})\sqrt{Q_-}-\widetilde{h}_\perp(q^2)(m_{\lb}-m_{\lc})\sqrt{Q_+}\Big] \, , \nn \\[4pt]
	H^{T-L,-1/2}_{+1/2,0,+1}=& \, \, -C_{T_L}\frac{\sqrt{2}}{\sqrt{q^2}}\Big[h_\perp(q^2)(m_{\lb}+m_{\lc})\sqrt{Q_-}+\widetilde{h}_\perp(q^2)(m_{\lb}-m_{\lc})\sqrt{Q_+}\Big],  \nn \\[4pt]
	H^{T-L,+1/2}_{-1/2,0,-1}=& \, \, C_{T_L}\frac{\sqrt{2}}{\sqrt{q^2}}\Big[h_\perp(q^2)(m_{\lb}+m_{\lc})\sqrt{Q_-}-\widetilde{h}_\perp(q^2)(m_{\lb}-m_{\lc})\sqrt{Q_+}\Big],  \nn \\[10pt]
	H^{T-L,+1/2}_{+1/2,+1,-1}=& \, \, -C_{T_L}\Big[h_+(q^2)\sqrt{Q_-}+\widetilde{h}_+(q^2)\sqrt{Q_+}\Big], \nn \\[4pt]
	H^{T-L,-1/2}_{-1/2,+1,-1}=& \, \, -C_{T_L}\Big[h_+(q^2)\sqrt{Q_-}-\widetilde{h}_+(q^2)\sqrt{Q_+}\Big] \, .
   \end{align}
and,
   \begin{align}
	H^{T-R,-1/2}_{-1/2,t,0}=& \, \, C_{T_R}\Big[h_+(q^2)\sqrt{Q_-}+\widetilde{h}_+(q^2)\sqrt{Q_+}\Big], \nn \\[10pt]
	H^{T-R,+1/2}_{+1/2,t,0}=& \, \, C_{T_R}\Big[h_+(q^2)\sqrt{Q_-}-\widetilde{h}_+(q^2)\sqrt{Q_+}\Big], \nn \\[10pt]
	H^{T-R,-1/2}_{+1/2,t,+1}=& \, \, -C_{T_R}\frac{\sqrt{2}}{\sqrt{q^2}}\Big[h_\perp(q^2)(m_{\lb}+m_{\lc})\sqrt{Q_-}-\widetilde{h}_\perp(q^2)(m_{\lb}-m_{\lc})\sqrt{Q_+}\Big],\nn \\[10pt]
	H^{T-R,+1/2}_{-1/2,t,-1}=& \, \, -C_{T_R}\frac{\sqrt{2}}{\sqrt{q^2}}\Big[h_\perp(q^2)(m_{\lb}+m_{\lc})\sqrt{Q_-}+\widetilde{h}_\perp(q^2)(m_{\lb}-m_{\lc})\sqrt{Q_+}\Big],\nn \\[10pt]
	H^{T-R,-1/2}_{+1/2,0,+1}=& \, \, C_{T_R}\frac{\sqrt{2}}{\sqrt{q^2}}\Big[h_\perp(q^2)(m_{\lb}+m_{\lc})\sqrt{Q_-}-\widetilde{h}_\perp(q^2)(m_{\lb}-m_{\lc})\sqrt{Q_+}\Big],\nn \\[10pt]
	H^{T-R,+1/2}_{-1/2,0,-1}=& \, \, -C_{T_R}\frac{\sqrt{2}}{\sqrt{q^2}}\Big[h_\perp(q^2)(m_{\lb}+m_{\lc})\sqrt{Q_-}+\widetilde{h}_\perp(q^2)(m_{\lb}-m_{\lc})\sqrt{Q_+}\Big],\nn \\[10pt]
	H^{T-R,+1/2}_{+1/2,+1,-1}=& \, \, C_{T_R}\Big[h_+(q^2)\sqrt{Q_-}-\widetilde{h}_+(q^2)\sqrt{Q_+}\Big],\nn \\[10pt]
	H^{T-R,-1/2}_{-1/2,+1,-1}=& \, \,
	C_{T_R}\Big[h_+(q^2)\sqrt{Q_-} + \widetilde{h}_+(q^2)\sqrt{Q_+}\Big].
\end{align}
The other non-vanishing helicity amplitudes of tensor type are related to the above by 
\begin{align}
	H^{T-L(R), \, \lambda_{\lb}}_{\lambda_{\lc}, \, \lambda, \, \lambda^\prime}=-H^{T-L(R),\, \lambda_{\lb}}_{\lambda_{\lc}, \, \lambda^\prime,\lambda} \, .
\end{align}
\subsection{Kinematics in leptonic frame and spinor representation}
The polarisation vectors associated with the virtual gauge boson in the $W^{*}$ rest frame, are written as:
\begin{equation}\label{eq:polWst}
	\epsilon^{\mu }(t) = \left(1; 0, 0, 0\right) \, ,\qquad  
	\epsilon^{\mu}(\pm 1) = \frac{1}{\sqrt{2}}\left(0;\pm 1, -i, 0\right) \, , \qquad 
	\epsilon^{\mu}(0) = \left(0; 0, 0, -1\right).
\end{equation}
The three-momentum and energy of the $\ell$ lepton in this frame can be written as
\begin{equation}
	|\mathbf{p}_{\ell}|=|\mathbf{p}_{\chi}|= \frac{\lambda^{1/2} (q^2, \, m_{\ell}^2, \, m_{\chi}^2) }{2 \sqrt{q^2}}, \qquad
	E_\ell= \frac{q^2+m_\ell^2-m_{\chi}^2}{2 \sqrt{q^2}}, \qquad  
	E_\chi =\frac{q^2+m_{\chi}^2-m_\ell^2}{2 \sqrt{q^2}}.
\end{equation}
The lepton spinors with three momentum $p_{\ell}$ along any arbitrary direction $p_{\ell}(\theta_\ell, \phi=0)$ with polar angle $\theta_\ell$, and azimuthal angle $\phi$ are calculated as follows
\begin{eqnarray}
	\label{spinor3}
	\bar{u}_\ell(+{\textstyle \frac{1}{2}},p_\ell)&=& \sqrt{E_\ell+m_\ell}\left( \cos (\theta_\ell/2),~~\sin (\theta_\ell/2),~~~
	\frac{-|\mathbf{p}_{\ell}|}{E_\ell+m_\ell}\cos (\theta_\ell/2),~~~\frac{-|\mathbf{p}_{\ell}|}{E_\ell+m_\ell}\sin (\theta_\ell/2) \right) \,,\nonumber \\
	\bar{u}_\ell(-{\textstyle \frac{1}{2}},p_{\ell})&=& \sqrt{E_\ell+m_\ell}\left( -\sin (\theta_\ell/2),~~~\cos (\theta_\ell/2),
	~~~\frac{-|\mathbf{p}_\ell|}{E_\ell+m_\ell}\sin (\theta_\ell/2),~~~\frac{|\mathbf{p}_{\ell}|}{E_\ell+m_\ell}\cos (\theta_\ell/2) \right) \,,\nonumber \\
	v_{\bar{X}_\ell}({\textstyle \frac{1}{2}},p_{\bar{X}_\ell}) &=& \sqrt{E_X+m_X}
	\left(\begin{array}{c} \frac{|\mathbf{p}_{\ell}|}{E_X+m_X}\cos (\theta_\ell/2) ~~~ \frac{|\mathbf{p}_{\ell}|}{E_X+m_X} \sin (\theta_\ell/2) ~~~ -\cos (\theta_\ell/2) ~~~ -\sin (\theta_\ell/2)  \end{array}\right)^T \,, \nonumber \\
	v_{\bar{X}_\ell}({\textstyle -\frac{1}{2}},p_{\bar{X}_\ell}) &=& \sqrt{E_X+m_X}
	\left(\begin{array}{c} \frac{|\mathbf{p}_{\ell}|}{E_X+m_X} \sin (\theta_\ell/2) ~~~-\frac{|\mathbf{p}_\ell|}{E_X+m_X}\cos (\theta_\ell/2)  ~~~ \sin (\theta_\ell/2) ~~~ -\cos (\theta_\ell/2)  \end{array}\right)^T \, .
\end{eqnarray}
\subsection{Leptonic Helicity Amplitudes}\label{subsec:leptonic_helicity}
In this section, we provide the explicit mathematical expressions of the leptonic helicity amplitudes defined in eq.~\eqref{eq:leptonic_helicity_amp}. We will write the expression of leptonic amplitude for including the dark-fermions, from which we can get the amplitudes for massless neutrinos by setting $ m_{\chi} \to 0$. In the calculation of the lepton helicity amplitudes, we worked in the rest frame of the virtual vector boson, the $\ell-\chi$ rest frame. We have defined the angle $\theta_\ell$ as the angle between the three-momenta of the $\ell$ and the $\Lambda_c$ in this frame. Using the above leptonic spinors and polarization vectors for the $W^*$ boson eq.~\eqref{eq:polWst}, we obtained the leptonic helicity amplitudes, which are presented in the following items:
\begin{itemize}
	\item The leptonic helicity amplitude associated with the vector current involving a left-handed massive neutrino (dark-fermion) is calculated as follows:
	\begin{subequations}
		\begin{eqnarray}\label{eq:leptonic_vector_left}
			L^{+\frac{1}{2}, \, VA-L}_{+ \frac{1}{2}} & = & \, \left\{ 1 \,, \,  \frac{\sin \theta_{\ell}}{\sqrt{2}}\, , \, -\frac{\sin \theta_{\ell}}{\sqrt{2}} \,, \, -\cos \theta_{\ell} \right\} X_{-+} \,, \\
			L^{+ \frac{1}{2}, VA-L}_{- \frac{1}{2}} & = & \, \left\{ 0 \,, \, -\frac{(1-\cos \theta_{\ell} )}{\sqrt{2}} \, , \, -\frac{(1+\cos \theta_{\ell} )}{\sqrt{2}}   \,, \, \sin \theta_{\ell} \right\} X_{--} \,, \\
			L^{-\frac{1}{2}, VA-L}_{+ \frac{1}{2}} & = & \, \left\{ 0 \,, \, \frac{(1+\cos \theta_{\ell})}{\sqrt{2}}\,, \, \frac{(1-\cos \theta_{\ell})}{\sqrt{2}} \,, \, \sin \theta_{\ell} \right\} X_{++} \,, \\
			L^{-\frac{1}{2}, VA-L}_{-\frac{1}{2}} & = & \, \left\{ 1 \,,- \frac{\sin \theta_{\ell}}{\sqrt{2}}  \,,  \frac{\sin \theta_{\ell}}{\sqrt{2}} \,, \cos \theta_{\ell} \right\}X_{+-}\,.
		\end{eqnarray}
	\end{subequations}
\item The leptonic helicity amplitude associated with the vector current involving a right-handed massive neutrino is calculated as follows:
\begin{subequations}
	\begin{eqnarray}\label{eq:leptonic_vector_right}
		L^{+\frac{1}{2},VA-R}_{+\frac{1}{2}} & = & \, \left\{-1 \, , \,  + \frac{\sin \theta_{\ell}}{\sqrt{2}} \,, \, - \frac{\sin \theta_{\ell}}{\sqrt{2}} \,,\, -\cos \theta_{\ell} \right\} X_{+-} \,, \\
		L^{+\frac{1}{2},VA-R}_{-\frac{1}{2}} & = &  \left\{ 0 \,,\,  - \frac{(1-\cos \theta_{\ell})}{\sqrt{2}} \, ,\, - \frac{(1+\cos \theta_{\ell} )}{\sqrt{2}}   \,,\, \sin \theta_{\ell} \right\} X_{++}\,, \\
		L^{-\frac{1}{2},VA-R}_{+\frac{1}{2}} & = & \, \left\{ 0 \,,\,  \frac{(1+\cos \theta_{\ell})}{\sqrt{2}}\,, \,\frac{(1-\cos \theta_{\ell})}{\sqrt{2}} \,, \, \sin \theta_{\ell} \right\} X_{--} \,,\\
		L^{-\frac{1}{2}, VA-R}_{-\frac{1}{2}} & = & \left\{ -1 \,, \, -\frac{\sin \theta_{\ell}}{\sqrt{2}}  \,, \,  \frac{\sin \theta_{\ell}}{\sqrt{2}} \,, \cos \theta_{\ell} \right\}X_{-+}\,.
	\end{eqnarray}
\end{subequations}

\item The leptonic helicity amplitude associated with the scalar current involving a left-handed massive neutrino is calculated as follows:
\begin{subequations}
	\begin{eqnarray}
		&L^{+\frac{1}{2},  \, SP-L}_{+\frac{1}{2}}  = &  \, X_{++} \\
		& L^{+\frac{1}{2}, \, SP-L}_{-\frac{1}{2}} = & \, 0 \\
		& L^{-\frac{1}{2}, \, SP-L}_{+\frac{1}{2}} = & \, 0 \\
		& L^{-\frac{1}{2}, \, SP-L}_{-\frac{1}{2}}  = & \, X_{--}
	\end{eqnarray}
\end{subequations}
\item The leptonic helicity amplitude associated with the scalar current involving a right-handed massive neutrino is calculated as follows: 
\begin{subequations}
	\begin{eqnarray}
		& L^{+\frac{1}{2}, \, SP-R}_{\frac{1}{2}}  = & -\, X_{--} \\
		& L^{+\frac{1}{2}, \, SP-R}_{-\frac{1}{2}} = & 0 \\
		& L^{-\frac{1}{2}, \, SP-R}_{\frac{1}{2}} = & 0 \\
		& L^{-\frac{1}{2}, \, SP-R}_{-\frac{1}{2}}  = & - \, X_{++}
	\end{eqnarray}
\end{subequations}
\item The leptonic helicity amplitude associated with the tensor current involving a left-handed massive neutrino is calculated as follows: 
\begin{subequations}
	\begin{eqnarray}
		L^{+\frac{1}{2} \, , \, T-L}_{+\frac{1}{2},t \,, \, \lambda'} & = &  \left\{  0\,, + \frac{\sin \theta_{\ell}}{\sqrt{2}} \,, - \frac{\sin \theta_{\ell}}{\sqrt{2}} \,, -\cos \theta_{\ell} \right\} X_{++} \,,\\
		L^{+\frac{1}{2} \, , \, T-L}_{-\frac{1}{2},t \,, \, \lambda'} & = & \left\{  0\,, \, -\frac{(1-\cos \theta_{\ell})}{\sqrt{2}} \,, -\frac{(1+\cos \theta_{\ell})}{\sqrt{2}} \,, \sin \theta_{\ell} \right\} X_{+-} \,,\\
		L^{-\frac{1}{2} \, , \, T-L}_{\frac{1}{2},t \,, \, \lambda'}& = & \left\{  0\,, \, \frac{(1+\cos \theta_{\ell})}{\sqrt{2}} \,, \frac{(1-\cos \theta_{\ell})}{\sqrt{2}} \,, \sin \theta_{\ell} \right\} X_{-+} \,,\\
		L^{-\frac{1}{2} \, , \, T-L}_{-\frac{1}{2}\,t \,, \, \lambda'} & = &  \left\{  0\,, -\frac{\sin \theta}{\sqrt{2}} \,, \frac{\sin \theta_{\ell} }{\sqrt{2}} \,, \cos\theta_{\ell} \right\} X_{--} \,,
	\end{eqnarray}
	\begin{eqnarray}
		\nonumber \\
		L^{+\frac{1}{2} \, , \, T-L}_{+\frac{1}{2}, 0 \,, \, \lambda'}& = & \left\{  \cos \theta_{\ell} \,, -\frac{\sin \theta_{\ell}}{\sqrt{2}} \,, -\frac{\sin \theta_{\ell}}{\sqrt{2}}\,, 0\right\} X_{++} \,,\\
		L^{+\frac{1}{2} \, , \, T-L}_{-\frac{1}{2},0 \,, \, \lambda'}& = & \left\{ -\sin \theta_{\ell} \,, \frac{1-\cos \theta_{\ell}}{\sqrt{2}} \,, - \frac{(1+\cos \theta_{\ell})}{\sqrt{2}} \,, 0 \right\} X_{+-} \,,\\
		L^{-\frac{1}{2} \, , \, T-L}_{+\frac{1}{2},0 \,, \, \lambda'} & = & \left\{ - \sin \theta_{\ell} \,,  - \frac{(1+\cos \theta_{\ell}) }{\sqrt{2}} \,, + \frac{(1-\cos \theta_{\ell})}{\sqrt{2}} \,, 0 \right\} X_{-+} \,,\\
		L^{-\frac{1}{2} \, , \, T-L}_{-\frac{1}{2},0 \,, \, \lambda'} & = & - \left\{ -\cos \theta_{\ell} \,, \frac{\sin \theta_{\ell} }{\sqrt{2}} \,, \frac{\sin \theta_{\ell}}{\sqrt{2}} \,, 0 \right\} X_{--} \,,
	\end{eqnarray}
	\begin{eqnarray}
		L^{+\frac{1}{2} \, , \, T-L}_{+\frac{1}{2},+1 \,, \, \lambda'} & = & \left\{ -\frac{\sin \theta_{\ell}}{\sqrt{2}} \,, 0 \,, - \cos\theta_{\ell} \,, \frac{\sin \theta_{\ell} }{\sqrt{2}}\right\} X_{++} \,,\\
		L^{+\frac{1}{2} \, , \, T-L}_{-\frac{1}{2},+1 \,, \, \lambda'} & = &  \left\{ \frac{ (1-\cos \theta_{\ell})}{\sqrt{2}} \,, 0 \,,  \frac{\sin \theta_{\ell}}{\sqrt{2}} \,, - \frac{(1-\cos \theta_{\ell} )}{\sqrt{2}} \right\} X_{+-} \,,\\
		L^{-\frac{1}{2} \, , \, T-L}_{+\frac{1}{2},+1 \,, \, \lambda'} & = &  \left\{ -\frac{(1+\cos \theta_{\ell})}{\sqrt{2}} \,, 0\,, \sin \theta_{\ell} \,, \frac{(1+\cos \theta_{\ell})}{\sqrt{2}}   \right\} X_{-+} \,,\\
		L^{-\frac{1}{2} \, , \, T-L}_{-\frac{1}{2},+1 \,, \, \lambda'} & = &  \left\{ \frac{\sin \theta_{\ell}}{\sqrt{2}} \,, 0 \,,  \cos \theta_{\ell} \,, -\frac{\sin \theta_{\ell}}{\sqrt{2}} \right\} X_{--} \,,
	\end{eqnarray}
	\begin{eqnarray}
		L^{+\frac{1}{2} \, , \, T-L}_{+\frac{1}{2}, -1 \,, \, \lambda'} & = & \left\{ \frac{\sin \theta_{\ell}}{\sqrt{2}} \,, \cos \theta_{\ell} \,, 0 \,, \frac{\sin \theta_{\ell}}{\sqrt{2}}   \right\} X_{++} \,,\\
		L^{+\frac{1}{2} \, , \, T-L}_{-\frac{1}{2}, -1 \,, \, \lambda'} & = & \left\{ \frac{(1+\cos \theta_{\ell})}{\sqrt{2}} \,, - \sin \theta_{\ell} \,, 0 \,,  \frac{(1+\cos \theta_{\ell})}{\sqrt{2}}  \right\} X_{+-} \,,\\
		L^{-\frac{1}{2} \, , \, T-L}_{+\frac{1}{2}, -1 \,, \, \lambda'} & = & \left\{ -\frac{(1-\cos \theta_{\ell} )}{\sqrt{2}} \,, - \sin \theta_{\ell} \,, 0 \,, -\frac{(1-\cos \theta_{\ell})}{\sqrt{{2}}} \right\} X_{-+} \,,\\
		L^{-\frac{1}{2} \, , \, T-L}_{-\frac{1}{2}, -1 \,, \, \lambda'} & = & \left\{-\frac{\sin \theta_{\ell}}{\sqrt{2}} \,, - \cos \theta_{\ell} \,, 0 \,, - \frac{\sin \theta_{\ell}}{\sqrt{2}}  \right\} X_{--} \,.
	\end{eqnarray}
\end{subequations}
\item The leptonic helicity amplitude associated with the tensor current involving a right-handed massive neutrino is calculated as follows: 
\begin{subequations}
	\begin{eqnarray}
		L^{+\frac{1}{2} \, , \, T-R}_{+\frac{1}{2}, t \,, \, \lambda'} & = &  \left\{ 0 \,, \frac{\sin \theta_{\ell}}{\sqrt{2}} \,, -\frac{\sin \theta_{\ell} }{\sqrt{2}}\,, -\cos \theta_{\ell} \right\} X_{--} \,,\\
		L^{+\frac{1}{2} \, , \, T-R}_{-\frac{1}{2}, t \,, \, \lambda'} & = &  \left\{ 0 \,, -\frac{(1-\cos \theta_{\ell})}{\sqrt{2}} \,, -\frac{(1+\cos \theta_{\ell} )}{\sqrt{2}} \,, \sin \theta_{\ell}  \right\} X_{-+} \,,\\
		L^{-\frac{1}{2} \, , \, T-R}_{+\frac{1}{2}, t \,, \, \lambda'} & = &  \left\{ 0\,, \frac{(1 + \cos \theta_{\ell} )}{\sqrt{2}}\,, \frac{(1-\cos \theta_{\ell})}{\sqrt{2}} \,, \sin \theta_{\ell}   \right\} X_{+-} \,,\\
		L^{-\frac{1}{2} \, , \, T-R}_{-\frac{1}{2}, t \,, \, \lambda'} & = & \left\{  0\,, -\frac{\sin \theta_{\ell}}{\sqrt{2}} \, , \frac{\sin \theta_{\ell} }{\sqrt{2}} \,, \cos\theta_{\ell} \right\} X_{++} \,,
	\end{eqnarray}
\end{subequations}
\begin{subequations}
	\begin{eqnarray}
		L^{+\frac{1}{2} \, , \, T-R}_{+\frac{1}{2},0 \,, \, \lambda'} & = &  \left\{ \cos \theta_{\ell} \,, \frac{\sin \theta_{\ell}}{\sqrt{2}} \,, \frac{\sin \theta_{\ell}}{\sqrt{2}} \,, 0  \right\} X_{--} \,,\\
		L^{+\frac{1}{2} \, , \, T-R}_{-\frac{1}{2},0 \,, \, \lambda'} & = &   \left\{  -\sin \theta_{\ell} \,,  -\frac{(1-\cos \theta_{\ell})}{\sqrt{2}} \,, \frac{(1+\cos \theta_\ell)}{\sqrt{2}} \,, 0 \right\} X_{-+} \,,\\
		L^{-\frac{1}{2} \, , \, T-R}_{+\frac{1}{2},0 \,, \, \lambda'} & = &   \left\{  -\sin \theta_{\ell} \,, \frac{(1+\cos \theta_{\ell})}{\sqrt{2}} \,,  -\frac{(1-\cos \theta_{\ell} )}{\sqrt{2}} \,, 0   \right\} X_{+-} \,,\\
		L^{-\frac{1}{2} \, , \, T-R}_{-\frac{1}{2},0 \,, \, \lambda'} & = &  \left\{ -\cos \theta_{\ell} \,, -\frac{\sin \theta_{\ell} }{\sqrt{2}} \,, -\frac{\sin \theta_{\ell}}{\sqrt{2}} \,, 0\right\} X_{++} \,,
	\end{eqnarray}
\end{subequations}
\begin{subequations}
	\begin{eqnarray}
		L^{+\frac{1}{2} \, , \, T-R}_{+\frac{1}{2},+1 \,, \, \lambda'} & = & \left\{ -\frac{\sin \theta_{\ell}}{\sqrt{2}} \,, 0\,, \cos \theta_{\ell} \,,\, -\frac{\sin \theta_{\ell}}{\sqrt{2}} \right\} X_{--} \,,\\
		L^{+\frac{1}{2} \, , \, T-R}_{-\frac{1}{2},+1 \,, \, \lambda'} & =&   \left\{ \frac{(1-\cos \theta_{\ell})}{\sqrt{2}} \,, 0 \,, -\sin \theta_{\ell}  \,,  \frac{(1-\cos \theta_{\ell})}{\sqrt{2}}  \right\} X_{-+} \,,\\
		L^{-\frac{1}{2} \, , \, T-R}_{+\frac{1}{2},+1 \,, \, \lambda'} & = &  \left\{ -\frac{(1+\cos \theta_{\ell})}{\sqrt{2}} \,, 0 \,, - \sin \theta_{\ell}  \,,  -\frac{(1+\cos \theta_{\ell})}{\sqrt{2}} \right\} X_{+-} \,,\\
		L^{-\frac{1}{2} \, , \, T-R}_{-\frac{1}{2},+1 \,, \, \lambda'} & = &  \left\{ \frac{\sin \theta_{\ell}}{\sqrt{2}} \,, 0 \,, - \cos \theta_{\ell}  \,, \frac{\sin \theta_{\ell}}{\sqrt{2}} \right\} X_{++} \,,
	\end{eqnarray}
\end{subequations}
\begin{subequations}
	\begin{eqnarray}
		L^{+\frac{1}{2} \, , \, TR}_{+\frac{1}{2},-1 \,, \, \lambda'} & = &  \left\{ \frac{\sin \theta_{\ell}}{\sqrt{2}} \,, - \frac{\cos \theta_{\ell}}{\sqrt{2}} \,, 0 \,,  -\frac{\sin \theta_{\ell}}{\sqrt{2}}  \right\} X_{--} \,,\\
		L^{+\frac{1}{2} \, , \, TR}_{-\frac{1}{2}, -1 \,, \, \lambda'} & = &  \left\{ \frac{(1+\cos \theta_{\ell})}{\sqrt{2}} \,, \frac{\sin \theta_{\ell}}{\sqrt{2}} \,, 0 \,, -\frac{(1+\cos \theta_{\ell})}{\sqrt{2}}   \right\} X_{-+} \,,\\
		L^{-\frac{1}{2} \, , \, TR}_{+\frac{1}{2},-1 \,, \, \lambda'} & =&   \left\{ -\frac{(1-\cos \theta_{\ell})}{\sqrt{2}} \,, \frac{\sin \theta_{\ell}}{\sqrt{2}} \,, 0 \,, \frac{(1-\cos \theta_{\ell})}{\sqrt{2}}\right\} X_{+-} \,,\\
		L^{-\frac{1}{2} \, , \, TR}_{-\frac{1}{2},-1 \,, \, \lambda'} & = &  \left\{ - \frac{\sin \theta_{\ell} }{\sqrt{2}} \,, \cos \theta_{\ell} \,, 0 \,, \frac{\sin \theta_{\ell}}{\sqrt{2}} \right\} X_{++} \,.
	\end{eqnarray}
\end{subequations}
\end{itemize}
The leptonic tensor current will change sign under the exchange of polarisation index, i.e.,
\begin{equation}\label{eq:leptonic_tensor_relation}
	L_{\lambda_{\chi}, \, \lambda, \, \lambda'}^{\lambda_{\ell}, \, T-L(R)} = - L_{\lambda_{\chi}, \, \lambda', \, \lambda}^{\lambda_{\ell}, \, T-L(R)} \,,
\end{equation}
where, $\lambda, \lambda'=\{ t, +1, -1, 0 \}$, and the abbreviation we defined
	\begin{eqnarray}\label{eq:kinetic_func}
		X_{\pm \pm } & = & \frac{\bigl(E_{\ell}+m_{\ell} \pm |\mathbf{p}_\ell|\bigr) \bigl(E_{X}+m_{X} \pm |\mathbf{p}_\ell|\bigr)}{ \ \sqrt{E_{\ell}+m_{\ell}} \,\sqrt{E_{X}+m_{X}}}  \,. \nonumber \\ 
\end{eqnarray}
\section{Four-Fold Decay Distribution}
\label{sec:fourfold_decayDist}
We consider $\Lb \to \Lc \ell \chi$ followed by the decay $\Lc(p_{\lc}, \epsilon) \to \Lambda (p_\Lambda) \pi(p_\pi)$. The differential decay width of the process $\Lb \to \lc( \to \Lambda \pi) \ell \bar{\chi}$ can be written as:
\begin{align}~\label{eq:fivefold_decaywidth}
		\frac{d}{ dk^2}\left(\frac{d^4 \Gamma }{d q^2 \, d\cos \theta_{\ell} \, d\cos \theta_{\Lambda} \, d\phi} \right) = & \frac{\lambda^{1/2}(p_{\lb}^2, p_{\lc}^2, q^2)}{1024 (2\pi)^6 m_{\lb}^3} 
		\frac{\lambda^{1/2}(k^2, p_\Lambda^2, p_\pi^2)}{m_{k}^2}  
		\frac{\lambda^{1/2}(q^2, p_{\ell}^2, p_{\chi}^2)}{q^2} \nn \\ &  \times \sum_{\lambda, \lambda_{\ell}} \sum_{\lambda_{\Lc}} \left| \mathcal{M}^{\lambda}_{\lambda_{\Lc}, \lambda_{\ell}}(\lb \to \Lambda \pi \ell \bar{\chi}) \right|^2 \, ,
	\end{align}
where, $\lambda(a, b, c) = a^2 + b^2 + c^2 - 2 a b - 2 a c - 2 b c$ is the K{\"a}llen $\lambda$ function. In the above equation $\mathcal{M}_{\lambda_{\lc}, \lambda_{\ell}}^{\lambda}(\Lb \to \Lambda \pi \ell \bar{\chi})$ represents the four-body helicity amplitude and the parameters $\lambda_\ell$, $\lambda_{\lc}$ and $\lambda$, respectively, represent the helicities of the charged lepton, $\lc$ baryon and the off-shell W-boson.
 
We can write the invariant amplitude for the $\Lambda_{b} \to \Lambda_c^+(\to \Lambda \pi^+) \ell^- \, \bar{\chi}_{\ell}$ decay as:
\begin{equation}\label{eq:matrix_amp_four_body}
	\mathcal{M}_{\lambda_{\lc}, \lambda_{\ell}}^{\lambda}(\lb \to \Lambda \pi \ell \, \bar{\rm \chi}) = \sum_{\lambda_{\Lambda}} \mathcal{M}^{\lambda_{\lc}}_{ \lambda_{\Lambda}}(\lc \to  \Lambda \pi) \ \frac{i}{(k^2 - m_{\lc}^2) + i m_{\lc} \Gamma_{\lc} } \  
	\mathcal{M}_{\lambda_{\lc},\lambda_{\ell}}^{\lambda}(\lb \to \lc \ell \, \bar{\rm \chi})  
	\, .
\end{equation}
$\mathcal{M}^{\lambda_{\lc}}_{ \lambda_{\Lambda}}(\lc \to  \Lambda \pi)$ is the amplitude for the secondary decay. 
\vspace{0.5cm}
\noindent \paragraph{\underline{\rm \bf Secondary Decay $\Lambda_c^+\to \Lambda\pi^+$:}}
In the SM the decay $\Lambda_c^+  \to\Lambda \pi^+$ is described by the effective Hamiltonian
\begin{equation}\label{eq:heffds1}
	H^\text{eff}_{\Lc \to \Lambda \pi}=\frac{4 G_F}{\sqrt{2}} \, V_{ud}^* V_{us}
	\left[\bar{d} \gamma_\mu P_L u\right]\left[\bar{s} \gamma^\mu P_L c\right] \, .
\end{equation}
The hadronic matrix element which determines the $\Lambda_c \to \Lambda \pi$ decay can be parametrized as 
\begin{align}
	&
	\frac{4 G_F}{\sqrt{2}} \, V_{ud}^* V_{us} \bra{\Lambda(p_\Lambda, \lambda_{\Lambda}) \pi^+(p_\pi)} \left[\bar{d} \gamma_\mu P_L u\right]\left[\bar{s} \gamma^\mu P_L c\right]
	\ket{\Lc(p_{\Lc}, \lambda_{\Lambda_c})} \cr
	& = N_2 \big[\bar u(p_\Lambda, \lambda_\Lambda) \big(\xi \,\gamma_5 + \omega\big) u(p_{\Lc}, \lambda_{\Lambda_c})\big] \equiv 
	\mathcal{A}^{\lambda_{\Lambda_c}}_{\lambda_\Lambda} \, ,
\end{align}
with $N_2=\frac{4 G_F}{\sqrt{2}} \, V_{ud}^* V_{us}$. Two independent hadronic parameters that parametrize this hadronic transition, which we denote as $\xi$ and $\omega$. In terms of the kinematic variables ($\xi$, $\omega$, $\theta_\Lambda$), the helicity amplitudes for the secondary decay can be obtained as
\begin{align}
	\mathcal{A}^{+1/2}_{+1/2} &= N_2  \left(\sqrt{r_+} \, \omega - \sqrt{r_-} \, \xi \right) \cos\frac{\theta_\Lambda}{2} \,, 
	\cr
	\mathcal{A}^{+1/2}_{-1/2} &= N_2  \left(\sqrt{r_+} \, \omega + \sqrt{r_-} \, \xi \right) \sin\frac{\theta_\Lambda}{2}\, 
	e^{i \phi}  \,, \cr
	\mathcal{A}^{-1/2}_{+1/2} &= N_2  \left(-\sqrt{r_+} \, \omega + \sqrt{r_-} \, \xi \right) \sin\frac{\theta_\Lambda}{2}\,
	e^{-i \phi} \,, \cr
	\mathcal{A}^{-1/2}_{-1/2} &= N_2  \left(\sqrt{r_+}\, \omega + \sqrt{r_-} \,\xi \right) \cos\frac{\theta_\Lambda}{2} \, ,
\end{align}
where we abbreviate,
\begin{equation}
	r_{\pm} \equiv (m_{\Lambda_c} \pm m_{\Lambda})^2 - m_{\pi}^2 \, .
\end{equation}
The corresponding helicity contributions to the decay width can be defined as
\begin{equation}
	\Gamma_2(\lambda_{\Lambda_c}^{(a)},\lambda_{\Lambda_c}^{(b)}) = \frac{\sqrt{r_+ r_-}}{16 \pi m_{\Lambda_c}^3} \,
	\sum_{\lambda_\Lambda} \mathcal{A}^{\lambda_{\Lambda_c}^{(a)}}_{\lambda_\Lambda} \, \bigl(\mathcal{A}^{\lambda_{\Lambda_c}^{(b)}}_{\lambda_\Lambda}\bigr)^* \, ,
\end{equation}
which yield
\begin{align}
	\Gamma_2(+1/2, +1/2) &=  (1 + \alpha  \, \cos\theta_\Lambda) \, \Gamma(\Lambda_c) \, ,\qquad
	\Gamma_2(+1/2, -1/2) = -\alpha \, \sin\theta_\Lambda \, e^{i \phi} \, \Gamma(\Lambda_c) \, ,\cr
	\Gamma_2(-1/2, -1/2) &= (1 - \alpha \, \cos\theta_\Lambda) \,\Gamma(\Lambda_c) \,,\qquad
	\Gamma_2(-1/2, +1/2) = -\alpha \, \sin\theta_\Lambda e^{-i \phi}\, \Gamma(\Lambda_c) \, .
\end{align}
Here the $\Lambda^+_c \to \Lambda \pi^+$ decay width is given as
\begin{align}
	\Gamma(\Lambda_c)
	= \frac{|N_2|^2 \sqrt{r_+ r_-}}{16 \pi m_{\Lambda_c}^3} \left(
	r_- \,|\xi|^2 + r_+ \, |\omega|^2 \right) \, .
\end{align}
and the parity-violating decay parameter $\alpha_\Lambda$ reads
\begin{equation}
	\alpha_\Lambda = \frac{-2 \,Re[\omega \,\xi]}{\sqrt{\frac{r_-}{r_+}} \, |\xi|^2 + \sqrt{\frac{r_+}{r_-}} \, |\omega|^2} = \alpha^{\text{exp}}_\Lambda \, .
\end{equation}
Then the total decay width is evaluated as:
\begin{equation}
	\Gamma(\lc \to \Lambda \pi) = \frac{\sqrt{r_+ r_-}} {16 \pi m_{\Lc}^3}\bigl|\mathcal{A}^{\lambda_{\lc}}_{\lambda_\Lambda}\bigr|^2 \, .
\end{equation}
\noindent \paragraph{\underline{\bf Total Decay Distribution:}}
The square of the total amplitude is given by
\small{
	\begin{equation}\label{eq:LbLc_diffDecaywidth}
		\left| \mathcal{M}_{\lambda_{\lc}, \lambda_{\ell}}^{\lambda}(\lb \to \Lambda \pi \ell \bar{\chi}) \right|^2 =   \frac{\pi}{m_{\lc} \Gamma_{\lc}} \delta(k^2 - m_{\lc}^2) \biggl| \sum_{\lambda_{\Lambda}} \mathcal{M}_{\lambda_{\lc},\lambda_{\ell}}^{\lambda}(\lb \to \lc \ell \bar{\chi})   \mathcal{M}^{\lambda_{\lc}}_{ \lambda_{\Lambda}}(\lc \to \Lambda \pi)  \biggr|^2 \, .
	\end{equation} 
}
The origin of the Dirac-delta function in this amplitude square is the square of the modulus of the Breit-Wigner propagator of the $\lc$ baryon. In the present case, the decay width $ \Gamma_{\lc} << m_{\lc} $, in narrow width approximation, we can write this as 
\begin{equation}
	\frac{1}{(k^2 - m_{\lc}^2)^2 + m_{\lc}^2 \Gamma_{\lc}^2} \quad \xrightarrow{\text{\small$\Gamma_{\lc} << m_{\lc}$} }{}\quad  \frac{\pi}{m_{\lc} \Gamma_{\lc}} \delta(k^2 - m_{\lc}^2) \, .
\end{equation}  
After integrating eq.~\eqref{eq:fivefold_decaywidth} with respect to $dk^2$, we obtain the following four-fold angular decay distribution:
\begin{align}\label{eq:fourfold_decaywidth}
	\frac{d^4 \Gamma }{d q^2 \, d\cos \theta_{\Lambda} \,  d\cos \theta_{\ell} \, d\phi} &= \frac{\lambda^{1/2}(m_{\lb}^2, m_{\lc}^2, q^2)}{1024 \, (2 \pi)^6 m_{\lb}^3} \frac{\lambda^{1/2}(m_{\Lc}^2, m_{\Lambda}^2, m_{\pi}^2)}{m_{\Lc}^2}  \frac{\lambda^{1/2}(q^2, m_{\ell}^2, m_{\rm X}^2)}{q^2}  \frac{\pi}{m_{\lc} \Gamma_{\lc}} \times \nonumber \\ 
	&\qquad \sum_{\lambda, \, \lambda_{\ell}, \, \lambda_{\Lambda}}  \biggl| \sum_{\lambda_{\lc}} \mathcal{M}_{\lambda_{\lc},\lambda_{\ell}}^{\lambda}( \Lb \to \lc \ell \bar{\chi})  \mathcal{M}^{\lambda_{\lc}}_{ \lambda_{\Lambda}}(\lc \to \Lambda \pi)  \biggr|^2 \, .
\end{align}
To obtain the above equation, we have used the following integration
\begin{equation}
	\int \frac{1}{m_{\lc} k}\delta(k^2 - m_{\lc}^2) dk^2 \rightarrow \frac{1}{m_{\lc} ^2} \, .
\end{equation}
The eq.~\eqref{eq:fourfold_decaywidth} can be reduced to 
\begin{equation}\label{eq:diffrate1}
	\frac{d^4 \Gamma}{dq^2\, d \cos\theta_\ell\,	d\cos\theta_\Lambda\, d\phi} = \mathcal{N} \sum_{\lambda_{\ell}, \lambda} \sum_{ \lambda_{\lc}}
	\Bigl|\mathcal{M}_{\lambda_{\lc},\lambda_{\ell}}^{\lambda} \, \mathcal{M}^{\lambda_{\lc}}_{ \lambda_{\Lambda}}\Bigr|^2 \, .
\end{equation}
We have obtained the above expression of eq.~\eqref{eq:diffrate1} using the narrow width approximation for the Breit-Wigner propagator. The helicity amplitudes $\mathcal{M}_{\lambda_{\lc},\lambda_{\ell}}^{\lambda}$ for $\lc \to \Lambda \pi^+$ decay are also discussed earlier in this section. With the different pieces collected together we have the following expression for $\mathcal{N}$:  
\begin{equation} 	
	\mathcal{N} =  \frac{1}{2^{16} \pi^5 m_{\lb}^3 m_{\lc}^3 \Gamma_{\lc} } \; \frac{ \sqrt{\lambda(m_{\lb}^2, m_{\lc}^2, q^2)}
		\sqrt{\lambda(m_{\lc}^2, m_{\Lambda}^2, m_{\pi}^2)}
		\sqrt{\lambda(q^2, m_\ell^2, m_X^2)}} {q^2}  \, \, \,  .
\end{equation}
\section{Angular Distribution of $\Lambda_b^0 \to \Lambda_c^+(\to \Lambda \pi^+) \ell^- X_{\rm inv}$ Decay}~\label{appsec:angular_Obs}
In this section, we discuss the angular distribution of the four-body final state for the decay process $\lb\to \lc(\to \Lambda \pi) \ell \chi$. This analysis allows us to construct a multitude of observables. We consider the transition process $\lb(p_{\Lb}) \to \lc(p_{\Lc}) \ell(p_\ell) \chi(p_\chi)$, where $\lc(p_{\Lc})$ decays to $\Lambda(p_{\Lambda})$ and $\pi(p_{\pi})$ on the mass shell. We calculate the full decay distribution by expanding the right-hand side of eq.~\eqref{eq:diffrate1} by performing the summation over the helicities $\lambda_{\Lambda}$ and $\lambda_{\lc}$. Upon summing over $\lambda_{\Lambda}$ and $\lambda_{\Lc}$, the dependence on the secondary decay angle $\theta_\Lambda$ emerges accompanying the secondary branching fraction. The resulting four-fold decay distribution is then obtained as follows:
\begin{align}~\label{eq:totl_dist}
	\frac{d^4 \Gamma (\Lambda_b \to \Lambda_c(\to \Lambda \pi) \ell \rm \, X_{\rm inv})}{dq^2 \, d \cos\theta_\ell \, d\cos\theta_\Lambda \, d\phi}
    &=\frac{2^4 \pi m^3_{\lc} \mathcal{N} }{\sqrt{r_+ r_-}}  \mathcal{B}(\lc \to \Lambda \pi) \times \nn \\
    &\quad  \sum_{\lambda, \lambda_{\ell}}
    \Biggl[\mathcal{M}_{\frac{1}{2}, \lambda_{\ell}}^{\lambda} \bigl(\mathcal{M}_{\frac{1}{2}, \lambda_{\ell}}^{\lambda}\bigr)^*
    (1+\alpha \cos \theta_{\Lambda}) 
    + \mathcal{M}_{\frac{1}{2},\lambda_{\ell}}^{\lambda} \bigl(\mathcal{M}_{-\frac{1}{2}, \lambda_{\ell}}^{\lambda}\bigr)^*
    (-\alpha \sin \theta_{\Lambda} e^{i \phi}) \nn \\
    &	+ \mathcal{M}_{-\frac{1}{2},\lambda_{\ell}}^{\lambda} \bigl(\mathcal{M}_{-\frac{1}{2}, \lambda_{\ell}}^{\lambda}\bigr)^*
    (1-\alpha \cos \theta_{\Lambda})
    + \mathcal{M}_{-\frac{1}{2}, \lambda_{\ell}}^{\lambda} \bigl(\mathcal{M}_{\frac{1}{2}, \lambda_{\ell}}^{\lambda}\bigr)^*
    (-\alpha \sin \theta_{\Lambda} e^{-i \phi}) \Biggr]	\nn \\
    & = \frac{3}{8\pi}K(q^2, \cos\theta_\ell, \cos\theta_\Lambda, \phi) \, .
\end{align}
The general expression of $K$ is given by:
\begin{align}
	K(q^2, \cos\theta_\ell, \cos\theta_\Lambda, \phi)
	& =
	\big( K_{1ss} \sin^2\theta_\ell +\, K_{1cc} \cos^2\theta_\ell + K_{1c} \cos\theta_\ell\big) \,\cr
	&  + \big( K_{2ss} \sin^2\theta_\ell +\, K_{2cc} \cos^2\theta_\ell + K_{2c} \cos\theta_\ell\big) \cos\theta_\Lambda
	\cr
	&  + \big( K_{3sc}\sin\theta_\ell \cos\theta_\ell + K_{3s} \sin\theta_\ell\big) \sin\theta_\Lambda \cos\phi\cr
	&  + \big( K_{4sc}\sin\theta_\ell \cos\theta_\ell + K_{4s} \sin\theta_\ell\big) \sin\theta_\Lambda \sin\phi \, .
\end{align}
These angular coefficients $K_i$s are related to the hadronic helicity amplitudes and, consequently, to the transition form factors. 
Including the contributions from a neutral massive fermion alongside the SM rate, the total four-fold differential decay distribution can be written as
\begin{equation}~\label{eq:total_dist}
	\frac{d^4 \Gamma (\Lambda_b \to \Lambda_c(\to \Lambda \pi) \ell \rm \, \bar{X}_{\mathrm{inv}})}{dq^2 \, d \cos\theta_\ell \, d\cos\theta_\Lambda \, d\phi} 
	= \frac{d^4 \Gamma (\Lambda_b \to \Lambda_c \ell \bar{\nu})}{dq^2 \, d \cos\theta_\ell \, d\cos\theta_\Lambda \, d\phi} 
	+ \frac{d^4 \Gamma (\Lambda_b \to \Lambda_c \ell \rm \, \bar{\chi})}{dq^2 \, d \cos\theta_\ell \, d\cos\theta_{\Lambda} \, d\phi} \, .
\end{equation}
The corresponding angular coefficients can be extracted as
\begin{equation}\label{eq:angular_coefficients}
	K_i(q^2, m_{\ell}^2, m_{X_{\mathrm{inv}}}^2) 
	= K_i(q^2, m_{\ell}^2)^{\Lambda_b \to \Lambda_c \ell \bar{\nu} } 
	+ K_i(q^2, m_{\ell}^2, m_{\chi}^2)^{ \Lambda_b \to \Lambda_c \ell \rm \, \bar{\chi}} \qquad .
\end{equation}
The normalized angular distribution, in which systematic uncertainties cancel in the ratio, is defined as
\begin{equation}
    \hat{K}_i(q^2)= \frac{K_i(q^2, m_{\ell}^2)^{\Lambda_b \to \Lambda_c \ell \bar{\nu} } 
	+ K_i(q^2, m_{\ell}^2, m_{\chi}^2)^{\Lambda_b \to \Lambda_c \ell \rm \, \bar{\chi}}}{\frac{d\Gamma(\Lambda_b \to \Lambda_c \ell \rm \, \bar{X}_{\mathrm{inv}})}{dq^2}} \, ,
    \label{eq:ang_normalised}
\end{equation}  
\paragraph{\underline{\rm \bf Other Angular Observables :}}
By integrating over the polar angle $\theta_\ell, \, \theta_{\Lambda}$ and the azimuthal angle $\phi$,  we can derive the differential decay rate distribution, which is dependent solely on the momentum transfer to the di-lepton, as follows:
\begin{align}\label{eq:dGammadq2_general}
   \frac{d\Gamma}{d \, q^2} &=2K_{1ss}(q^2)+K_{1cc}(q^2) \, .
\end{align}
This definition suggests that $2\hat{K}_{1ss}(q^2)+\hat{K}_{1cc}(q^2)=1$ \,. 

We have defined a couple of other important observables which we have presented in the following enumerated items: 
\begin{enumerate}
\item Among the other important observables, the forward-backward asymmetry concerning the leptonic scattering angle, normalized to the differential rate, is defined as
	\begin{equation}
		A^\ell_\text{FB}(q^2) = \frac{3}{2} \, \frac{K_{1c}(q^2)}{2 K_{1ss}(q^2) + K_{1cc}(q^2)} \, ,   
	\end{equation}
	which we can obtain from eq.~\ref{eq:totl_dist}, following the integration as given below
	\begin{equation}
		A^\ell_\text{FB}(q^2) =\frac{\int_{0}^{2\pi} d\phi\int_{-1}^{1} d(\cos \theta_{\Lambda}) \big[\int_{0}^{1}-\int_{-1}^{0}\big]d(\cos\theta_{\ell}) K(q^2, \cos\theta_\ell, \cos\theta_\Lambda, \phi)}{d \Gamma/dq^2} \, .
	\end{equation}
	
	\item The analogous asymmetry for the baryonic scattering angle reads
	\begin{align}
		A^{\Lambda_c}_\text{FB}(q^2)
		& = \frac{1}{2} \, \frac{2 K_{2ss}(q^2) + K_{2cc}(q^2)}{2 K_{1ss}(q^2) + K_{1cc}(q^2)} \, . 
	\end{align}
	Using eq.~\ref{eq:totl_dist} we can define the above asymmetry as  
	\begin{equation}
		A^{\Lambda_c}_\text{FB}(q^2)  =\frac{\int_{0}^{2\pi} d\phi \int_{-1}^{1} d(\cos \theta_{\ell})\big[\int_{0}^{1}-\int_{-1}^{0}\big]d(\cos \theta_{\Lambda}) K(q^2, \cos\theta_\ell, \cos\theta_\Lambda, \phi)}{{d \Gamma/dq^2}} \, .
	\end{equation}
	
	\item 	For $\Lambda_b\to \Lambda^+_c (\to \Lambda \pi^+) \ell^-\bar{\chi}$ decays, one could also define a combined forward-backward asymmetry 
	\begin{align}
		A^{\Lambda_c \ell}_\text{FB}(q^2) & = \frac{3}{4} \, \frac{K_{2c}(q^2)}{2 K_{1ss}(q^2) + K_{1cc}(q^2)} \, ,
	\end{align}
	which we have obtained following the integrations as given below
	
	\begin{align}
		A^{\Lambda_c \ell }_\text{FB}(q^2) & =\frac{\int_{0}^{2\pi} d\phi \big[\int_{0}^{1}-\int_{-1}^{0}\big]d(\cos \theta_{\Lambda})  I_\ell(q^2, \cos\theta_\Lambda, \phi) }{d \Gamma/dq^2} \, ,
	\end{align}
	with 
	\begin{equation}
		I_\ell(q^2, \cos\theta_\Lambda, \phi) = \Big[\int_{0}^{1}   -\int_{-1}^{0}  \Big]d(\cos\theta_{\ell})  K(q^2, \cos\theta_\ell, \cos\theta_\Lambda, \phi) \, .
	\end{equation}
	Here, $I_{\ell}(q^2, \cos\theta_\Lambda, \phi)$ is defined as the difference of the three-fold decay rates of a lepton moving in the forward and backward regions respectively. 
    \end{enumerate}
\section{Form Factor Shape}~\label{appsec:FormFactor_info}
The helicity amplitudes of the hadronic current depend on the non-perturbative form factors and their $q^2$ shapes. The lattice QCD results on the relevant form factors are available in ref.~\cite{Detmold:2015aaa, Datta:2017aue}. To obtain the $q^2$ shapes of the form factors we used $z$-parametrization, read as:
\begin{equation}
	f(q^2)=\frac{1}{1-q^2/(m_{\text{pole}}^f)^2}\Big{[}a_{0}+a_{1}z^{f}(q^2) + a_{2}[z^{f}(q^2)]^2\Big{]} \, ,
	\label{eq:fitness}
\end{equation}
where the parameters $a^f_{0}$ and $a^f_{1}$, $a^f_{2}$ are expansion coefficients. The expansion parameter $z^f$ is defined as
\begin{equation}
	z^f(q^2)=\frac{\sqrt{t_+^f-q^2}-\sqrt{t_+^f-t_0}}{\sqrt{t_+^f-q^2}+\sqrt{t_+^f-t_0}} \, ,
\end{equation}
with $t_+^f=(m_\text{pole}^f)^2$ and $t_{0}^f=(m_{\Lb} - m_{\Lc})^2$. The pole masses $m_{\text{pole}}^f$, used in eq.~\ref{eq:fitness} for the respective form factors, are listed together with their associated spin quantum numbers in table~\ref{tab:BCLpolmass}. The values of the parameters $a^f_{0}$, $a^f_{1}$ and $a^f_{2}$, together with their covariance matrix are provided in ref.~\cite{Detmold:2015aaa, Datta:2017aue} and are summarized in table~\ref{tab:LQCDffinfo}. We use these inputs to obtain the shapes of the form factors. Using these shapes, we have obtained the decay rate distributions and predicted many other related observables.  
\begin{table}[t]
	\renewcommand{\arraystretch}{1.2}
	\centering
	\setlength\tabcolsep{25pt}
	\begin{tabular}{*{4}{c}}
		\hline \hline
		\text{Form factor} &\text{$J^P$} & \text{Resonance} &\text{$m_{\text{pole}}^f$} \text{[GeV]}\\
		\hline
		\text{$f_0$}& \text{$0^+$} & \text{$B_{c}$} & $6.725$~\cite{Detmold:2015aaa}\\
		\text{$f_+$, $f_{\perp}$} & \text{$1^-$} & \text{$B^*_{c}$} & $6.332$ \cite{Detmold:2015aaa}\\
		\text{$g_0$}& \text{$0^-$} & \text{$B_{c}$} & $6.276$~\cite{Detmold:2015aaa} \\
		\text{$g_+$, $g_{\perp}$} & \text{$1^+$} & \text{$B_{c}$} & $6.768~$\cite{Detmold:2015aaa}\\
		\text{$h_+$, $h_{\perp}$} & \text{$1^-$} & \text{$B^*_{c}$} & $6.332$~\cite{Datta:2017aue}\\
		\text{$\tilde{h}_+$, $\tilde{h}_{\perp}$} & \text{$1^+$} & \text{$B_{c}$} & $6.768$~\cite{Datta:2017aue}\\
		\hline
	\end{tabular}
	\caption{Pole masses used in the $z$ parametrization.}
	\label{tab:BCLpolmass}
\end{table}
\begin{table}
\rotatebox{90}{
	\renewcommand*{\arraystretch}{1.6}
	\resizebox{1.3 \textwidth}{!}{
	\begin{tabular}{|c|c||*{28}{c}|}
		\hline
  Parameters & Results 	& $a_0^{f_+}$ & $a_1^{f_+}$ & $a_2^{f_+}$ & $a_0^{f_0}$ \
	& $a_1^{f_0}$ & $a_2^{f_0}$ & $a_0^{f_\perp}$ & $a_1^{f_\perp}$ \
	&$a_2^{f_\perp}$ & $a_0^{g_\perp, g_+}$ & $a_1^{g_+}$ & $a_2^{g_+}$  \
	&$a_0^{g_0}$ & $a_1^{g_0}$ & $a_2^{g_0}$ & $a_1^{g_\perp}$ & $a_2^{g_\perp}$ \
	& $a_0^{h_+}$ & $a_1^{h_+}$ & $a_2^{h_+}$ & $a_0^{h_\perp}$ & $a_1^{h_\perp}$ \
	& $a_2^{h_\perp}$ & $a_0^{\tilde{h}_{\perp}, \tilde{h}_{+}}$ & $a_1^{\tilde{h}_+}$ & $a_2^{\tilde{h}_+}$ & $a_1^{\tilde{h}_{\perp}}$ & $a_2^{\tilde{h}_{\perp}}$ \\
	\hline \hline
	$a_0^{f_+}$ & $0.810 \pm 0.028$ & $1.0$  &  $-0.522$  &  $0.162$  &  $0.712$  &  $-0.266$  &  \
	$-0.029$  &  $0.626$  &  $-0.268$  &  $0.008$  &  $0.199$  &  \
	$-0.131$  &  $-0.028$  &  $0.284$  &  $-0.184$  &  $0.044$  &  \
	$-0.161$  &  $0.009$  &  $0.185$  &  $-0.127$  &  $0.$  &  $0.179$  & \
	$-0.156$  &  $-0.013$  &  $0.086$  &  $-0.126$  &  $0.003$  &  \
	$-0.109$  &  $-0.001$  \\
	$a_1^{f_+}$& $-4.748 \pm 0.943$  & $0.$  &  $1.0$  &  $-0.66$  &  $-0.32$  &  $0.428$  &  $0.065$  &  \
	$-0.255$  &  $0.262$  &  $-0.01$  &  $-0.14$  &  $0.188$  &  $0.041$  \
	&  $-0.158$  &  $0.193$  &  $-0.036$  &  $0.17$  &  $0.003$  &  \
	$-0.137$  &  $0.191$  &  $0.01$  &  $-0.125$  &  $0.211$  &  $0.031$  \
	&  $-0.06$  &  $0.134$  &  $-0.006$  &  $0.134$  &  $-0.002$  \\
	$a_2^{f_+}$ & $0.786 \pm 8.802$ & $0.$  &  $0.$  &  $1.0$  &  $-0.035$  &  $0.131$  &  $0.094$  &  \
	$0.018$  &  $-0.015$  &  $0.03$  &  $-0.011$  &  $0.019$  &  $0.001$  \
	&$0.$  &  $-0.004$  &  $0.019$  &  $0.025$  &  $0.009$  &  $0.014$  \
	&$-0.01$  &  $0.003$  &  $0.009$  &  $-0.017$  &  $0.029$  &  \
	$-0.004$  &  $0.018$  &  $0.003$  &  $0.018$  &  $0.002$  \\
	$a_0^{f_0}$ & $0.739 \pm 0.023$ & $0.$  &  $0.$  &  $0.$  &  $1.0$  &  $-0.514$  &  $0.112$  &  \
	$0.519$  &  $-0.204$  &  $-0.001$  &  $0.254$  &  $-0.21$  &  $0.013$ \
	& $0.202$  &  $-0.148$  &  $0.01$  &  $-0.206$  &  $-0.008$  &  \
	$0.105$  &  $-0.097$  &  $-0.002$  &  $0.128$  &  $-0.134$  &  \
	$-0.023$  &  $0.119$  &  $-0.168$  &  $-0.003$  &  $-0.172$  &  \
	$0.005$  \\
	$a_1^{f_0}$ & $-4.563 \pm 0.943$ &$0.$  &  $0.$  &  $0.$  &  $0.$  &  $1.0$  &  $-0.524$  &  $-0.179$  \
	&  $0.228$  &  $0.009$  &  $-0.177$  &  $0.259$  &  $0.013$  &  \
	$-0.127$  &  $0.185$  &  $-0.009$  &  $0.234$  &  $0.013$  &  \
	$-0.093$  &  $0.176$  &  $0.014$  &  $-0.1$  &  $0.204$  &  $0.052$  & \
	$-0.08$  &  $0.187$  &  $-0.005$  &  $0.194$  &  $-0.011$  \\
	$a_2^{f_0}$ & $2.705 \pm 8.443$ & $0.$  &  $0.$  &  $0.$  &  $0.$  &  $0.$  &  $1.0$  &  $-0.022$  &  \
	$0.027$  &  $0.014$  &  $0.004$  &  $-0.015$  &  $0.03$  &  $-0.007$  \
	& $0.011$  &  $-0.003$  &  $-0.022$  &  $0.007$  &  $-0.017$  &  \
	$0.024$  &  $0.003$  &  $-0.009$  &  $0.007$  &  $0.019$  &  $0.003$  \
	&$-0.02$  &  $0.004$  &  $-0.019$  &  $0.014$  \\
	$a_0^{f_\perp}$& $1.094 \pm 0.044$ &$0.$  &  $0.$  &  $0.$  &  $0.$  &  $0.$  &  $0.$  &  $1.0$  &  \
	$-0.583$  &  $0.114$  &  $0.175$  &  $-0.125$  &  $-0.017$  &  \
	$0.288$  &  $-0.181$  &  $0.042$  &  $-0.132$  &  $-0.009$  &  \
	$0.201$  &  $-0.124$  &  $0.$  &  $0.18$  &  $-0.147$  &  $-0.011$  & \
	$0.071$  &  $-0.107$  &  $0.001$  &  $-0.086$  &  $-0.006$  \\
	$a_1^{f_\perp}$& $-6.441 \pm 1.501$ & $0.$  &  $0.$  &  $0.$  &  $0.$  &  $0.$  &  $0.$  &  $0.$  &  \
	$1.0$  &  $-0.465$  &  $-0.115$  &  $0.147$  &  $0.036$  &  \
	$-0.149$  &  $0.165$  &  $-0.034$  &  $0.132$  &  $0.01$  &  $-0.15$  \
	&  $0.208$  &  $0.011$  &  $-0.127$  &  $0.209$  &  $0.028$  &  \
	$-0.049$  &  $0.107$  &  $-0.004$  &  $0.102$  &  $0.004$  \\
	$a_2^{f_\perp}$ & $2.316 \pm 11.320$ & $0.$  &  $0.$  &  $0.$  &  $0.$  &  $0.$  &  $0.$  &  $0.$  &  $0.$ \
	&  $1.0$  &  $-0.001$  &  $0.$  &  $0.006$  &  $0.005$  &  $-0.006$  & \
	$0.009$  &  $-0.001$  &  $0.003$  &  $0.004$  &  $-0.006$  &  \
	$0.006$  &  $0.007$  &  $-0.02$  &  $0.025$  &  $0.002$  &  $-0.004$  \
	&  $0.001$  &  $-0.005$  &  $0.002$  \\
	$a_0^{g_\perp, g_+}$ & $0.684\pm 0.018 $ & $0.$  &  $0.$  &  $0.$  &  $0.$  &  $0.$  &  $0.$  &  $0.$  &  $0.$ \
	&  $0.$  &  $1.0$  &  $-0.444$  &  $0.088$  &  $0.707$  &  $-0.26$ \
	&  $0.013$  &  $-0.427$  &  $0.048$  &  $0.082$  &  $-0.063$  &  \
	$0.$  &  $0.107$  &  $-0.104$  &  $-0.007$  &  $0.119$  &  $-0.159$  & \
	$0.001$  &  $-0.164$  &  $0.009$  \\
	$a_1^{g_+}$ & $-4.379 \pm 0.695$ & $0.$  &  $0.$  &  $0.$  &  $0.$  &  $0.$  &  $0.$  &  $0.$  &  $0.$ \
	&  $0.$  &  $0.$  &  $1.0$  &  $-0.547$  &  $-0.279$  &  $0.344$  &  \
	$0.054$  &  $0.478$  &  $-0.138$  &  $-0.068$  &  $0.102$  &  $0.007$ \
	&  $-0.078$  &  $0.13$  &  $0.017$  &  $-0.09$  &  $0.186$  &  \
	$-0.005$  &  $0.197$  &  $-0.015$  \\
	$a_2^{g_+}$ & $1.281 \pm 7.365$ & $0.$  &  $0.$  &  $0.$  &  $0.$  &  $0.$  &  $0.$  &  $0.$  &  $0.$ \
	&  $0.$  &  $0.$  &  $0.$  &  $1.0$  &  $-0.045$  &  $0.119$  &  \
	$0.058$  &  $-0.148$  &  $0.269$  &  $-0.013$  &  $0.028$  &  $0.002$ \
	&  $-0.007$  &  $0.021$  &  $0.008$  &  $0.008$  &  $-0.009$  &  \
	$0.002$  &  $-0.01$  &  $0.019$  \\
	$a_0^{g_0}$ & $0.741 \pm 0.026$ & $0.$  &  $0.$  &  $0.$  &  $0.$  &  $0.$  &  $0.$  &  $0.$  &  $0.$ \
	&  $0.$  &  $0.$  &  $0.$  &  $0.$  &  $1.0$  &  $-0.552$  &  \
	$0.22$  &  $-0.302$  &  $0.006$  &  $0.141$  &  $-0.07$  &  $0.005$  & \
	$0.148$  &  $-0.114$  &  $0.001$  &  $0.089$  &  $-0.135$  &  \
	$0.002$  &  $-0.116$  &  $0.001$  \\
	$a_1^{g_0}$ & $-4.386 \pm 0.877$ & $0.$  &  $0.$  &  $0.$  &  $0.$  &  $0.$  &  $0.$  &  $0.$  &  $0.$ \
	& $0.$  &  $0.$  &  $0.$  &  $0.$  &  $0.$  &  $1.0$  &  $-0.769$  & \
	$0.244$  &  $0.019$  &  $-0.103$  &  $0.106$  &  $0.003$  &  \
	$-0.098$  &  $0.131$  &  $0.01$  &  $-0.059$  &  $0.129$  &  $-0.006$ \
	&  $0.125$  &  $-0.001$  \\
	$a_2^{g_0}$ & $1.338 \pm 8.044$ & $0.$  &  $0.$  &  $0.$  &  $0.$  &  $0.$  &  $0.$  &  $0.$  &  $0.$ \
	&  $0.$  &  $0.$  &  $0.$  &  $0.$  &  $0.$  &  $0.$  &  $1.0$  &  \
	$0.$  &  $0.041$  &  $0.029$  &  $-0.021$  &  $0.001$  &  $0.022$  &  \
	$-0.027$  &  $0.004$  &  $0.002$  &  $-0.003$  &  $0.004$  &  $0.002$ \
	&  $0.003$  \\
	$a_1^{g_\perp}$ & $-4.627 \pm 0.709$ & $0.$  &  $0.$  &  $0.$  &  $0.$  &  $0.$  &  $0.$  &  $0.$  &  $0.$ \
	&  $0.$  &  $0.$  &  $0.$  &  $0.$  &  $0.$  &  $0.$  &  $0.$  &  \
	$1.0$  &  $-0.503$  &  $-0.075$  &  $0.087$  &  $0.004$  &  $-0.086$  & \
	$0.118$  &  $0.013$  &  $-0.086$  &  $0.186$  &  $-0.005$  &  \
	$0.191$  &  $-0.017$  \\
	$a_2^{g_\perp}$ & $1.614 \pm 7.453$ &$0.$  &  $0.$  &  $0.$  &  $0.$  &  $0.$  &  $0.$  &  $0.$  &  $0.$ \
	&  $0.$  &  $0.$  &  $0.$  &  $0.$  &  $0.$  &  $0.$  &  $0.$  &  \
	$0.$  &  $1.0$  &  $-0.004$  &  $0.01$  &  $0.$  &  $-0.003$  &  \
	$0.008$  &  $0.004$  &  $-0.005$  &  $0.009$  &  $0.009$  &  $0.009$  \
	& $0.015$  \\
	$a_0^{h_+}$ & $0.967 \pm 0.057$ & $0.$  &  $0.$  &  $0.$  &  $0.$  &  $0.$  &  $0.$  &  $0.$  &  $0.$ \
	& $0.$  &  $0.$  &  $0.$  &  $0.$  &  $0.$  &  $0.$  &  $0.$  &  \
	$0.$  &  $0.$  &  $1.0$  &  $-0.506$  &  $0.067$  &  $0.848$  &  \
	$-0.347$  &  $-0.017$  &  $0.062$  &  $-0.099$  &  $0.003$  &  \
	$-0.084$  &  $-0.006$  \\
	$a_1^{h_+}$ & $-4.526 \pm 1.754$ & $0.$  &  $0.$  &  $0.$  &  $0.$  &  $0.$  &  $0.$  &  $0.$  &  $0.$ \
	&  $0.$  &  $0.$  &  $0.$  &  $0.$  &  $0.$  &  $0.$  &  $0.$  &  \
	$0.$  &  $0.$  &  $0.$  &  $1.0$  &  $-0.33$  &  $-0.234$  &  $0.336$  \
	&  $0.075$  &  $-0.043$  &  $0.12$  &  $-0.007$  &  $0.121$  &  \
	$0.005$  \\
	$a_2^{h_+}$ & $2.201 \pm 10.724$ & $0.$  &  $0.$  &  $0.$  &  $0.$  &  $0.$  &  $0.$  &  $0.$  &  $0.$ \
	&  $0.$  &  $0.$  &  $0.$  &  $0.$  &  $0.$  &  $0.$  &  $0.$  &  \
	$0.$  &  $0.$  &  $0.$  &  $0.$  &  $1.0$  &  $0.001$  &  $0.016$  &  \
	$0.022$  &  $0.$  &  $0.005$  &  $0.$  &  $0.006$  &  $0.002$  \\
	$a_0^{h_\perp}$ & $0.705 \pm 0.036$ & $0.$  &  $0.$  &  $0.$  &  $0.$  &  $0.$  &  $0.$  &  $0.$  &  $0.$ \
	&  $0.$  &  $0.$  &  $0.$  &  $0.$  &  $0.$  &  $0.$  &  $0.$  &  \
	$0.$  &  $0.$  &  $0.$  &  $0.$  &  $0.$  &  $1.0$  &  $-0.457$  &  \
	$0.077$  &  $0.083$  &  $-0.12$  &  $0.002$  &  $-0.108$  &  $-0.001$ \
	\\
	$a_1^{h_\perp}$ & $-4.105 \pm 0.839$ & $0.$  &  $0.$  &  $0.$  &  $0.$  &  $0.$  &  $0.$  &  $0.$  &  $0.$ \
	&  $0.$  &  $0.$  &  $0.$  &  $0.$  &  $0.$  &  $0.$  &  $0.$  &  \
	$0.$  &  $0.$  &  $0.$  &  $0.$  &  $0.$  &  $0.$  &  $1.0$  &  \
	$-0.501$  &  $-0.075$  &  $0.171$  &  $-0.009$  &  $0.172$  &  \
	$-0.006$  \\
	$a_2^{h_\perp}$ & $3.010 \pm 7.835$ & $0.$  &  $0.$  &  $0.$  &  $0.$  &  $0.$  &  $0.$  &  $0.$  &  $0.$ \
	&  $0.$  &  $0.$  &  $0.$  &  $0.$  &  $0.$  &  $0.$  &  $0.$  &  \
	$0.$  &  $0.$  &  $0.$  &  $0.$  &  $0.$  &  $0.$  &  $0.$  &  $1.0$  & \
	$-0.001$  &  $0.024$  &  $0.004$  &  $0.024$  &  $0.016$  \\
	$a_0^{\tilde{h}_{\perp}, \tilde{h}_{+}}$ & $0.676 \pm 0.033$ & $0.$  &  $0.$  &  $0.$  &  $0.$  &  $0.$  &  $0.$  &  $0.$  &  $0.$ \
	&  $0.$  &  $0.$  &  $0.$  &  $0.$  &  $0.$  &  $0.$  &  $0.$  &  \
	$0.$  &  $0.$  &  $0.$  &  $0.$  &  $0.$  &  $0.$  &  $0.$  &  $0.$  & \
	$1.0$  &  $-0.419$  &  $0.024$  &  $-0.441$  &  $0.056$  \\
	$a_1^{\tilde{h}_+}$ & $-4.363 \pm 0.751$ & $0.$  &  $0.$  &  $0.$  &  $0.$  &  $0.$  &  $0.$  &  $0.$  &  $0.$ \
	&  $0.$  &  $0.$  &  $0.$  &  $0.$  &  $0.$  &  $0.$  &  $0.$  &  \
	$0.$  &  $0.$  &  $0.$  &  $0.$  &  $0.$  &  $0.$  &  $0.$  &  $0.$  & \
	$0.$  &  $1.0$  &  $-0.481$  &  $0.498$  &  $-0.145$  \\
	$a_2^{\tilde{h}_+}$ & $2.274 \pm 8.077$ & $0.$  &  $0.$  &  $0.$  &  $0.$  &  $0.$  &  $0.$  &  $0.$  &  $0.$ \
	&  $0.$  &  $0.$  &  $0.$  &  $0.$  &  $0.$  &  $0.$  &  $0.$  &  \
	$0.$  &  $0.$  &  $0.$  &  $0.$  &  $0.$  &  $0.$  &  $0.$  &  $0.$  & \
	$0.$  &  $0.$  &  $1.0$  &  $-0.083$  &  $0.114$  \\
    $a_1^{\tilde{h}_{\perp}}$ & $-4.554 \pm 0.737$ & $0.$  &  $0.$  &  $0.$  &  $0.$  &  $0.$  &  $0.$  &  $0.$  &  $0.$ \
	&  $0.$  &  $0.$  &  $0.$  &  $0.$  &  $0.$  &  $0.$  &  $0.$  &  \
	$0.$  &  $0.$  &  $0.$  &  $0.$  &  $0.$  &  $0.$  &  $0.$  &  $0.$  & \
	$0.$  &  $0.$  &  $0.$  &  $1.0$  &  $-0.55$  \\
	$a_2^{\tilde{h}_{\perp}}$ & $3.085 \pm 7.904$ & $0.$  &  $0.$  &  $0.$  &  $0.$  &  $0.$  &  $0.$  &  $0.$  &  $0.$ \
	&  $0.$  &  $0.$  &  $0.$  &  $0.$  &  $0.$  &  $0.$  &  $0.$  &  \
	$0.$  &  $0.$  &  $0.$  &  $0.$  &  $0.$  &  $0.$  &  $0.$  &  $0.$  & \
	$0.$  &  $0.$  &  $0.$  &  $0.$  &  $1.0$  \\
	\hline
	\end{tabular}
}
}
\caption{$z$-expansion coefficients for the transition form factors, including central values, uncertainties, and correlations from the lattice QCD studies in refs.~\cite{Detmold:2015aaa, Datta:2017aue}.}
\label{tab:LQCDffinfo}
\end{table}
\section{$q^2$-Distribution of Angular Observables}~\label{appsec:angular_Obs2}
Here in this section, we present the $q^2$ distributions of the angular observables sensitive to $m_\chi$ for another representative benchmark point $m_\chi=0.5 \text{GeV}$ with WC $\lesssim \mathcal{O}(1)$.
\begin{figure*}[t]
	\centering
	\vspace{-0.1cm}
	\subfloat[]{
		\includegraphics[scale=0.40]{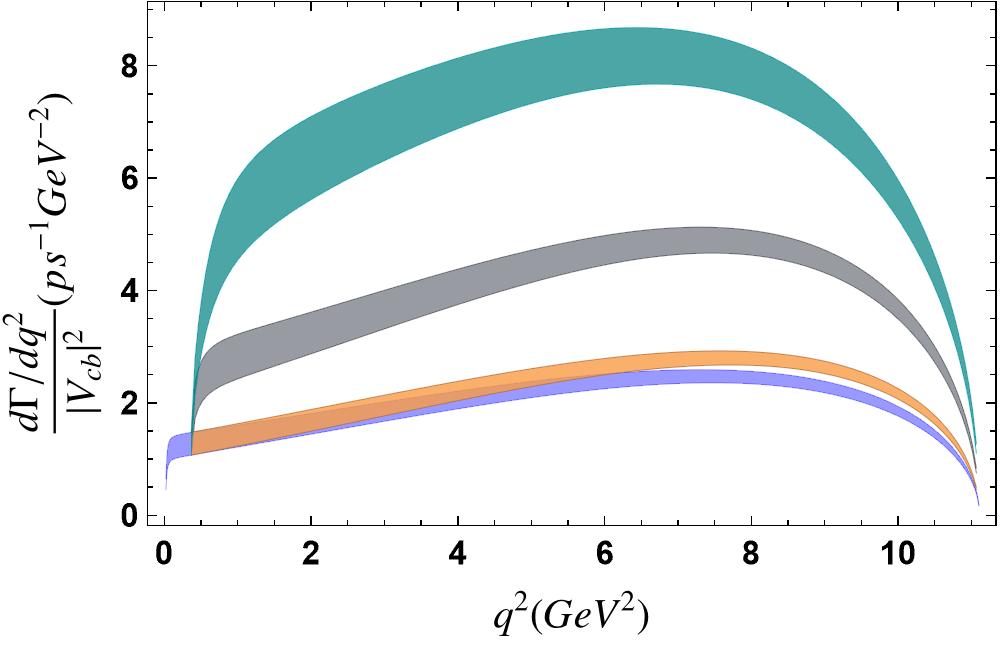}}
	\vspace{-0.1cm}\hspace{0.0002cm}
	\subfloat[]{\includegraphics[scale=0.40]{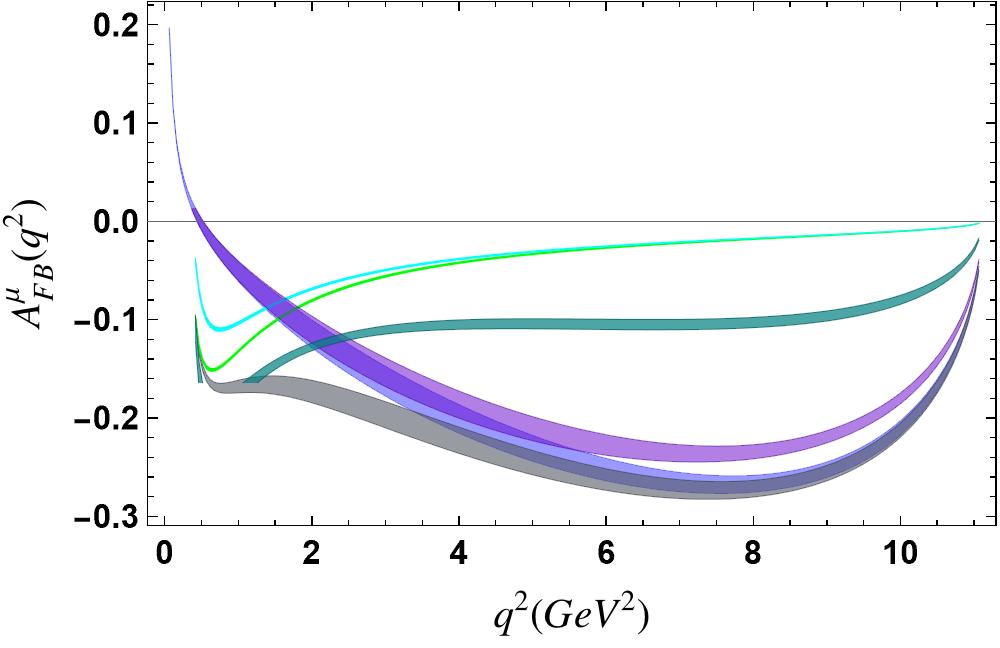}}
	\vspace{-0.1cm}\hspace{0.0002cm}
	\subfloat[]{\includegraphics[scale=0.40]{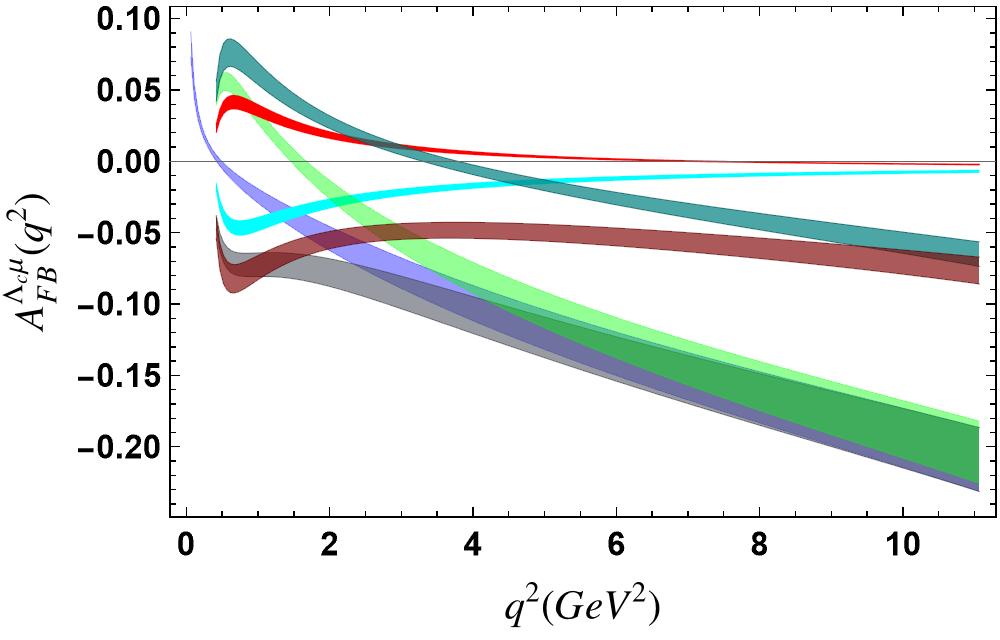}}
	\vspace{-0.1cm}\hspace{0.0002cm}
	\subfloat[]{\includegraphics[scale=0.40]{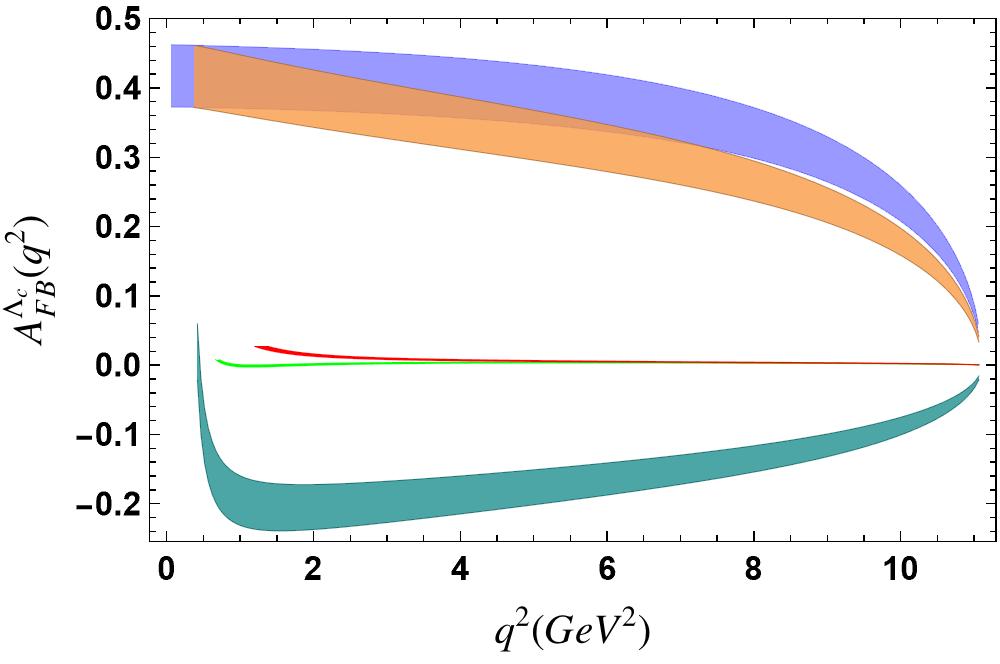}}
	\vspace{-0.1cm}\hspace{0.0002cm}
	\subfloat[]{\includegraphics[scale=0.40]{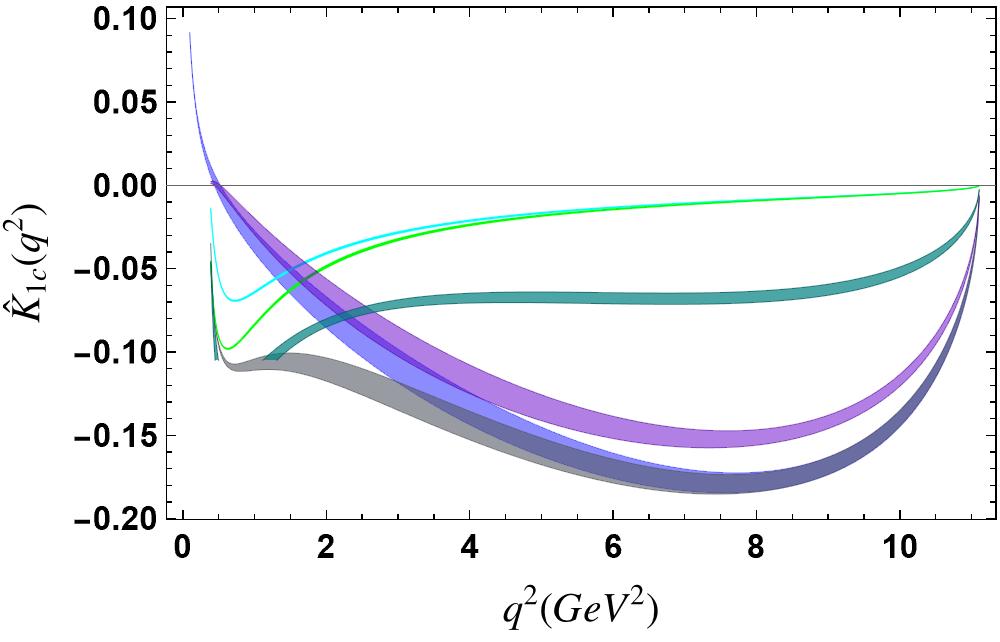}}
	\vspace{-0.1cm}\hspace{0.0003cm}
	\subfloat[]{\includegraphics[scale=0.40]{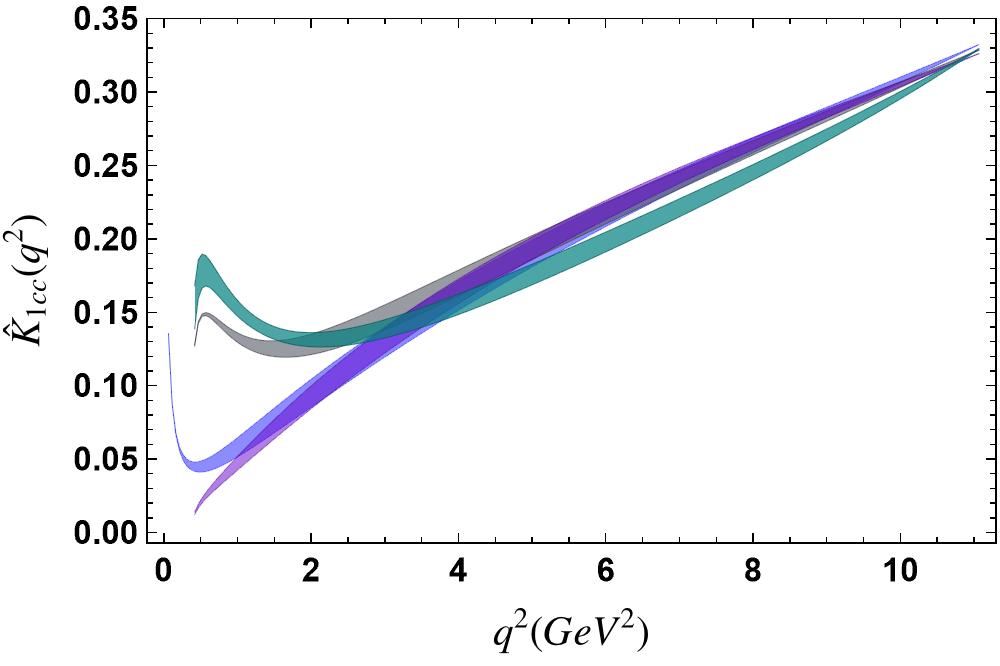}}
	\vspace{-0.1cm}\hspace{0.0002cm}
	\subfloat[]{\includegraphics[scale=0.40]{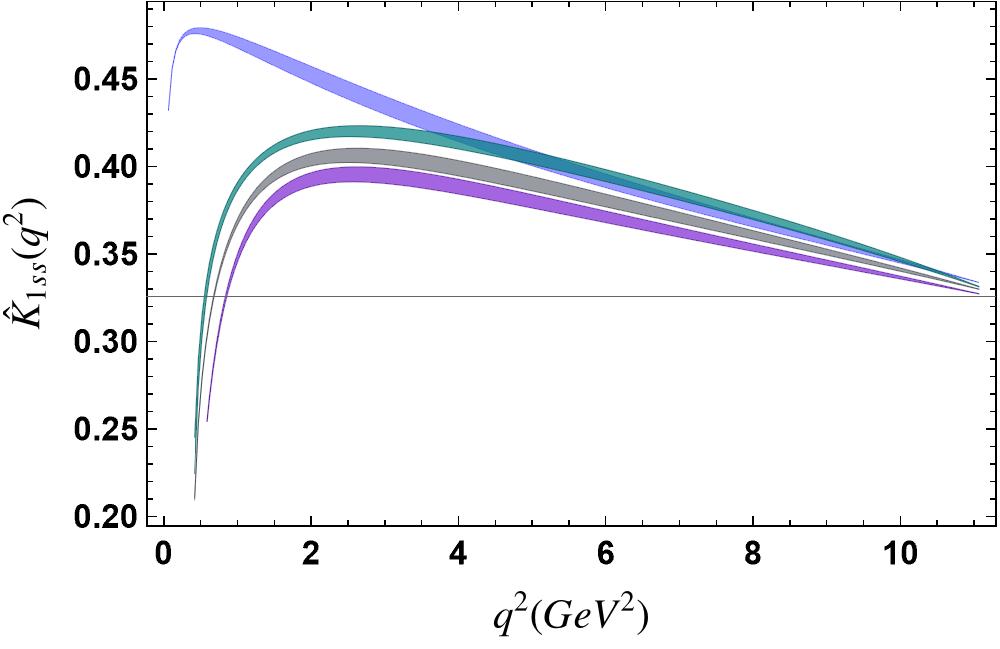}}
	\vspace{-0.1cm}\hspace{0.0002cm}
	\subfloat[]{\includegraphics[scale=0.40]{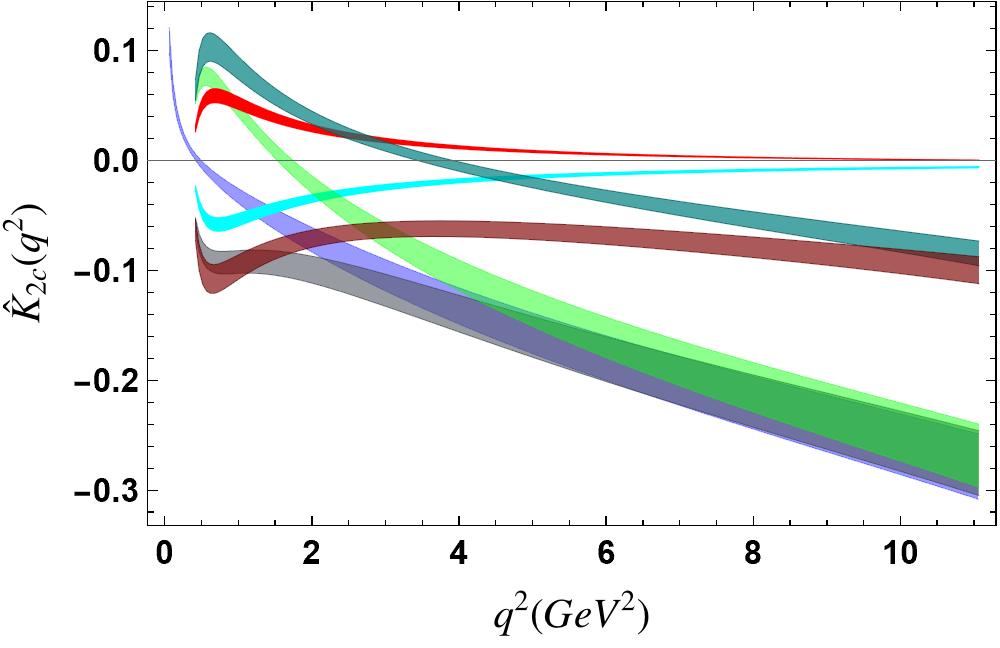}}
	\vspace{-0.1cm}\hspace{0.0002cm} 	
	\subfloat[]{\includegraphics[scale=0.40]{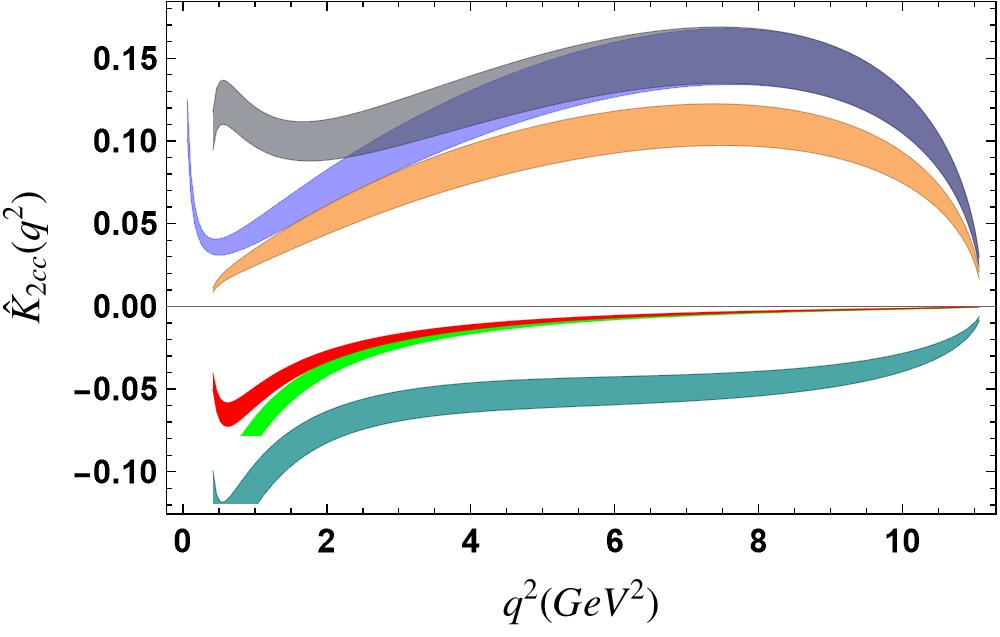}}
	\vspace{-0.1cm}\hspace{0.0002cm}
	\subfloat[]{\includegraphics[scale=0.40]{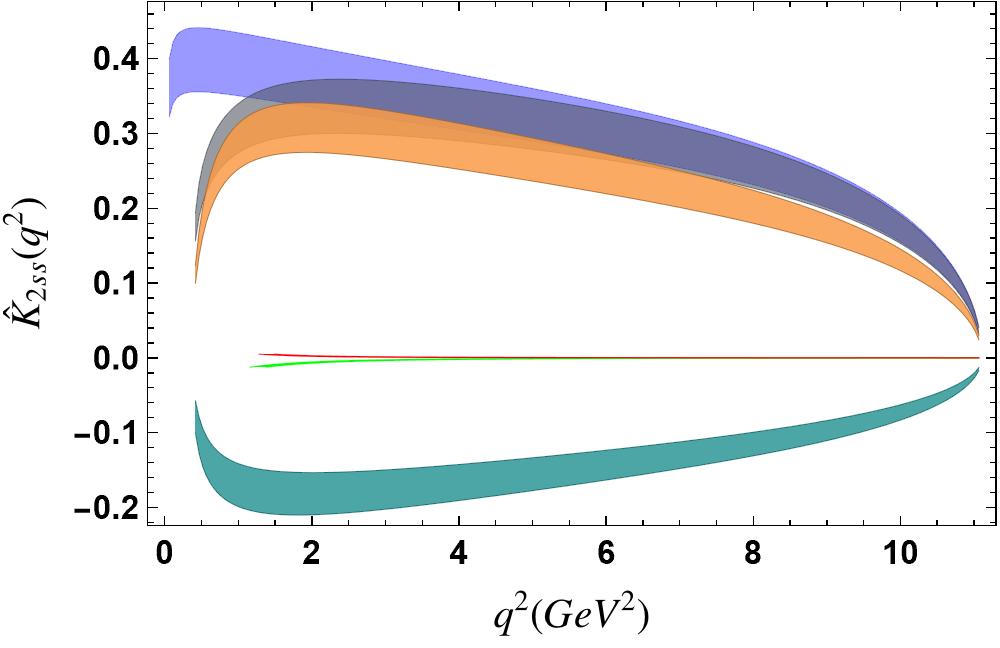}}
	\vspace{-0.1cm}\hspace{0.0002cm}
	\subfloat[]{\includegraphics[scale=0.40]{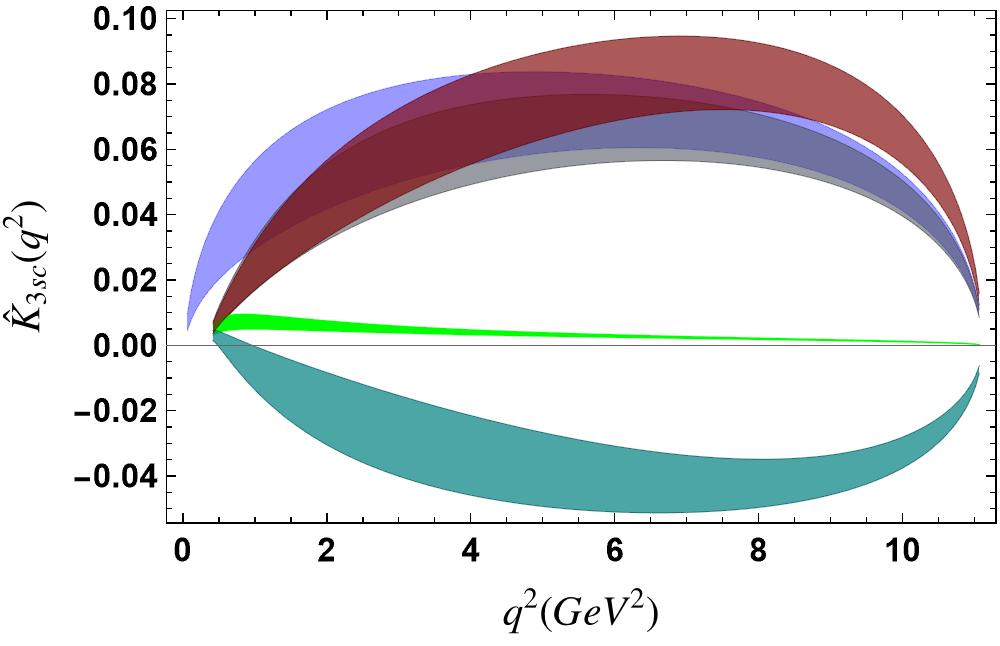}}
	\vspace{-0.1cm}\hspace{0.0002cm}
	\subfloat[]{\includegraphics[scale=0.40]{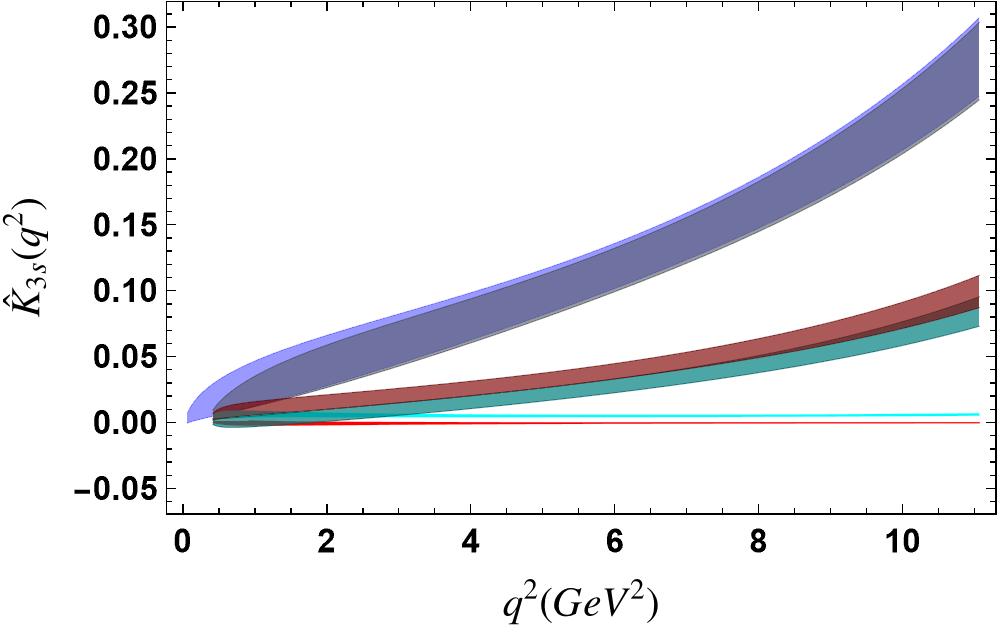}}
	\vspace{0.25cm}
	 \hspace{0.0002cm}{\includegraphics[scale=0.6]{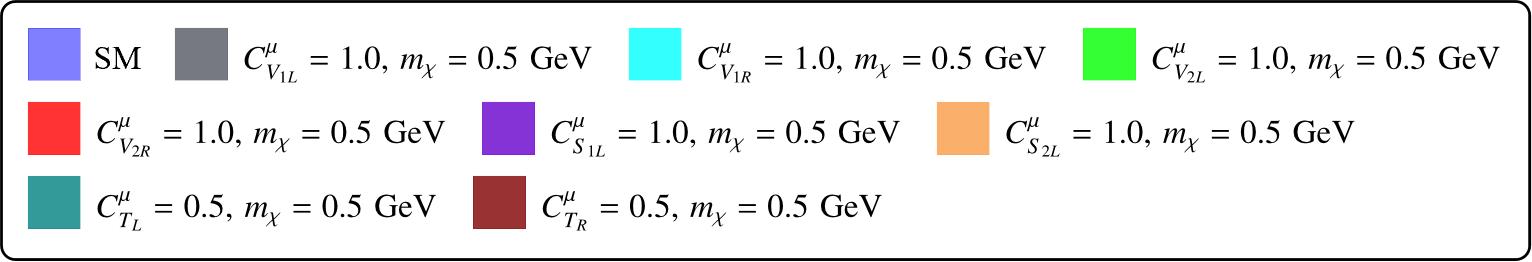}}
	\caption{The $q^2$-distribution of the angular observables in the one-operator scenarios, considering both left- and right-handed currents for benchmark values of WCs and $m_{\chi}$. For non-zero mass of the invisible particle, the region $q^2 \in [m_{\ell}^2, \,  (m_{\ell} + m_{\chi})^2]$ follows purely SM distribution, hence, not shown in the plots. The colored band shows the corresponding $1\sigma$ uncertainties of the observables.}
	\label{fig:Lc_fourfld_plots_mu0pt5}
\end{figure*}
\bibliographystyle{ieeetr}
\bibliography{lmblmc_RHN}
\end{document}